\input harvmac
% \draftmode

\overfullrule=0pt

\def\CP{{\cal P}}

\def\frac#1#2{{#1\over #2}}

\def\mut{ {\tilde \mu}}
\def\CR{{\cal R}}

\def\w{\omega}

\def\g{\gamma}

\def\ap{\alpha'}

%\def\NS{{\rm NS-NS}}
%\def\RR{{\rm R-R}}

%\MorrisBW
\lref\MorrisBW{ T.~R.~Morris, ``2-D Quantum Gravity, Multicritical
Matter And Complex Matrices,'' FERMILAB-PUB-90-136-T
%\href{http://www.slac.stanford.edu/spires/find/hep/www?r=fermilab-pub-90-136-t}{SPIRES entry}
}

\lref\GrossHE{ D.~J.~Gross and E.~Witten, ``Possible Third Order
Phase Transition In The Large N Lattice Gauge Theory,'' Phys.\
Rev.\ D {\bf 21}, 446 (1980).
%%CITATION = PHRVA,D21,446;%%
}

%\DiFrancescoRU
\lref\DiFrancescoRU{ P.~Di Francesco, ``Rectangular Matrix Models
and Combinatorics of Colored Graphs,'' Nucl.\ Phys.\ B {\bf 648},
461 (2003) [arXiv:cond-mat/0208037].
%%CITATION = COND-MAT 0208037;%%
}

%\MartinecHT
\lref\MartinecHT{ E.~J.~Martinec, G.~W.~Moore and N.~Seiberg,
``Boundary operators in 2-D gravity,'' Phys.\ Lett.\ B {\bf 263},
190 (1991).
%%CITATION = PHLTA,B263,190;%%
}

\lref\minahan{ J.~A.~Minahan,
``Matrix models with boundary terms and the generalized Painleve II equation,''
Phys.\ Lett.\ B {\bf 268}, 29 (1991);
%%CITATION = PHLTA,B268,29;%%
J.~A.~Minahan,
``Schwinger-Dyson equations for unitary matrix models with boundaries,''
Phys.\ Lett.\ B {\bf 265}, 382 (1991).
%%CITATION = PHLTA,B265,382;%%
}

\lref\PeriwalGF{ V.~Periwal and D.~Shevitz, ``Unitary Matrix
Models As Exactly Solvable String Theories,'' Phys.\ Rev.\ Lett.\
{\bf 64}, 1326 (1990);
%%CITATION = PRLTA,64,1326;%%
V.~Periwal and D.~Shevitz, ``Exactly Solvable
Unitary Matrix Models: Multicritical Potentials And
Correlations,'' Nucl.\ Phys.\ B {\bf 344}, 731 (1990).
%%CITATION = NUPHA,B344,731;%%
}

%\DouglasUP
\lref\DouglasUP{ M.~R.~Douglas, I.~R.~Klebanov, D.~Kutasov,
J.~Maldacena, E.~Martinec and N.~Seiberg, ``A new hat for the c =
1 matrix model,'' arXiv:hep-th/0307195.
%%CITATION = HEP-TH 0307195;%%
}

\lref\CrnkovicMS{ C.~Crnkovic, M.~R.~Douglas and G.~W.~Moore,
``Physical Solutions For Unitary Matrix Models,''
Nucl.\ Phys.\ B {\bf 360}, 507 (1991).
%%CITATION = NUPHA,B360,507;%%
}

%\DiFrancescoXZ
\lref\DiFrancescoXZ{ P.~Di Francesco, H.~Saleur and J.~B.~Zuber,
``Generalized Coulomb Gas Formalism For Two-Dimensional Critical
Models Based On SU(2) Coset Construction,'' Nucl.\ Phys.\ B {\bf
300}, 393 (1988).
%%CITATION = NUPHA,B300,393;%%
}

%\DijkgraafPP
\lref\DijkgraafPP{ R.~Dijkgraaf, S.~Gukov, V.~A.~Kazakov and
C.~Vafa, ``Perturbative analysis of gauged matrix models,''
arXiv:hep-th/0210238.
%%CITATION = HEP-TH 0210238;%%
}

\lref\Fukuda{ T.~Fukuda and K.~Hosomichi, ``Super Liouville theory
with boundary,'' Nucl.\ Phys.\ B {\bf 635}, 215 (2002)
[arXiv:hep-th/0202032].
%%CITATION = HEP-TH 0202032;%%
}

%\KostovXW
\lref\KostovXW{ I.~K.~Kostov, ``Solvable statistical models on a
random lattice,'' Nucl.\ Phys.\ Proc.\ Suppl.\  {\bf 45A}, 13
(1996) [arXiv:hep-th/9509124].
%%CITATION = HEP-TH 9509124;%%
}

%\HollowoodXQ
\lref\Hollowood{ T.~J.~Hollowood, L.~Miramontes, A.~Pasquinucci
and C.~Nappi, ``Hermitian versus anti-Hermitian one matrix models
and their hierarchies,'' Nucl.\ Phys.\ B {\bf 373}, 247 (1992)
[arXiv:hep-th/9109046].
%%CITATION = HEP-TH 9109046;%%
}

%\NappiBI
\lref\NappiBI{
C.~R.~Nappi,
``Painleve-II And Odd Polynomials,''
Mod.\ Phys.\ Lett.\ A {\bf 5}, 2773 (1990).
%%CITATION = MPLAE,A5,2773;%%
}

\lref\MartinecKA{ E.~J.~Martinec, ``The annular report on
non-critical string theory,'' arXiv:hep-th/0305148.
%%CITATION = HEP-TH 0305148;%%
}

\lref\AlexandrovNN{ S.~Y.~Alexandrov, V.~A.~Kazakov and
D.~Kutasov, ``Non-Perturbative Effects in Matrix Models and
D-branes,'' arXiv:hep-th/0306177.
%%CITATION = HEP-TH 0306177;%%
}

%\DalleyBR
\lref\DalleyBR{ S.~Dalley, C.~V.~Johnson, T.~R.~Morris and
A.~Watterstam, ``Unitary matrix models and 2-D quantum gravity,''
Mod.\ Phys.\ Lett.\ A {\bf 7}, 2753 (1992) [arXiv:hep-th/9206060].
%%CITATION = HEP-TH 9206060;%%
}

\lref\johnsonflows{ C.~V.~Johnson, T.~R.~Morris and A.~Watterstam,
``Global KdV flows and stable 2-D quantum gravity,'' Phys.\ Lett.\
B {\bf 291}, 11 (1992) [arXiv:hep-th/9205056].
%%CITATION = HEP-TH 9205056;%%
}

%\DalleyQG
\lref\DalleyQG{ S.~Dalley, C.~V.~Johnson and T.~Morris,
``Multicritical complex matrix models and nonperturbative 2-D
quantum gravity,'' Nucl.\ Phys.\ B {\bf 368}, 625 (1992).
%%CITATION = NUPHA,B368,625;%%
}

%\LafranceWY
\lref\LafranceWY{ R.~Lafrance and R.~C.~Myers, ``Flows For
Rectangular Matrix Models,'' Mod.\ Phys.\ Lett.\ A {\bf 9}, 101
(1994) [arXiv:hep-th/9308113].
%%CITATION = HEP-TH 9308113;%%
}

%\SeibergEB
\lref\SeibergEB{ N.~Seiberg, ``Notes On Quantum Liouville Theory
And Quantum Gravity,'' Prog.\ Theor.\ Phys.\ Suppl.\  {\bf 102},
319 (1990).
%%CITATION = PTPSA,102,319;%%
}

\lref\gelfand{ I.~M.~Gelfand and L.~A.~Dikii, ``Asymptotic
Behavior Of The Resolvent Of Sturm-Liouville Equations And The
Algebra Of The Korteweg-De Vries Equations,'' Russ.\ Math.\
Surveys {\bf 30}, 77 (1975) [Usp.\ Mat.\ Nauk {\bf 30}, 67
(1975)].
%%CITATION = RMSUA,30,77;%%
}

%\BrowerMN
\lref\BrowerMN{ R.~C.~Brower, N.~Deo, S.~Jain and C.~I.~Tan,
``Symmetry breaking in the double well Hermitian matrix models,''
Nucl.\ Phys.\ B {\bf 405}, 166 (1993) [arXiv:hep-th/9212127].
%%CITATION = HEP-TH 9212127;%%
}

%\CrnkovicWD
\lref\CrnkovicWD{ C.~Crnkovic, M.~R.~Douglas and G.~W.~Moore,
``Loop equations and the topological phase of multi-cut matrix
models,'' Int.\ J.\ Mod.\ Phys.\ A {\bf 7}, 7693 (1992)
[arXiv:hep-th/9108014].
%%CITATION = HEP-TH 9108014;%%
}

\lref\BerKleb{
M.~Bershadsky and I.~R.~Klebanov,
``Genus One Path Integral In Two-Dimensional Quantum Gravity,''
Phys.\ Rev.\ Lett.\  {\bf 65}, 3088 (1990).
%%CITATION = PRLTA,65,3088;%%
}
\lref\BerKlebnew{
M.~Bershadsky and I.~R.~Klebanov,
``Partition functions and physical states
in two-dimensional quantum gravity and supergravity,''
Nucl.\ Phys.\ B {\bf 360}, 559 (1991).
%%CITATION = NUPHA,B360,559;%%
}
\lref\igor{ I.~R.~Klebanov, ``String theory in two-dimensions,''
arXiv:hep-th/9108019.
%%CITATION = HEP-TH 9108019;%%
}

\lref\ginsparg{ P.~Ginsparg and G.~W.~Moore, ``Lectures On 2-D
Gravity And 2-D String Theory,'' arXiv:hep-th/9304011.
%%CITATION = HEP-TH 9304011;%%
}

%\DiFrancescoGinsparg
\lref\DiFrancescoGinsparg{P.~Di Francesco, P.~Ginsparg and
J.~Zinn-Justin,``2-D Gravity and random matrices,'' Phys.\ Rept.\
{\bf 254}, 1 (1995) [arXiv:hep-th/9306153].
%%CITATION = HEP-TH 9306153;%%
}

\lref\joe{J.~Polchinski, ``What is String Theory?''
arXiv:hep-th/9411028
%%CITATION = HEP-TH 9411028;%%
}

\lref\mgv{ J.~McGreevy and H.~Verlinde, ``Strings from tachyons:
The $c = 1$ matrix reloaded,'' arXiv:hep-th/0304224.
%%CITATION = HEP-TH 0304224;%%
}

%\SchomerusVV
\lref\SchomerusVV{ V.~Schomerus, ``Rolling tachyons from Liouville
theory,'' arXiv:hep-th/0306026.
%%CITATION = HEP-TH 0306026;%%
}

%\GaiottoYF
\lref\GaiottoYF{ D.~Gaiotto, N.~Itzhaki and L.~Rastelli, ``On the
BCFT description of holes in the c = 1 matrix model,''
arXiv:hep-th/0307221.
%%CITATION = HEP-TH 0307221;%%
}

%\GutperleIJ
\lref\GutperleIJ{ M.~Gutperle and P.~Kraus, ``D-brane dynamics in
the c = 1 matrix model,'' arXiv:hep-th/0308047.
%%CITATION = HEP-TH 0308047;%%
}

%\KapustinHI
\lref\KapustinHI{ A.~Kapustin, ``Noncritical superstrings in a
Ramond-Ramond background,'' arXiv:hep-th/0308119.
%%CITATION = HEP-TH 0308119;%%
}

%\GiveonWN
\lref\GiveonWN{ A.~Giveon, A.~Konechny, A.~Pakman and A.~Sever,
``Type 0 strings in a 2-d black hole,'' arXiv:hep-th/0309056.
%%CITATION = HEP-TH 0309056;%%
}

\lref\Harv{ J.L.~Karczmarek and A.~Strominger, ``Matrix
cosmology,'' arXiv:hep-th/0309138.
%%CITATION = HEP-TH 0309138;%%
}

\lref\Kitp{O.~DeWolfe, R.~Roiban, M.~Spradlin, A.~Volovich and
J.~Walcher, ``On the S-matrix of type 0 string theory,''
arXiv:hep-th/0309148.
%%CITATION = HEP-TH 0309148;%%
}

\lref\KlebanovKM{
I.~R.~Klebanov, J.~Maldacena and N.~Seiberg,
``D-brane decay in two-dimensional string theory,''
arXiv:hep-th/0305159.
%%CITATION = HEP-TH 0305159;%%
}

\lref\McGreevyEP{
J.~McGreevy, J.~Teschner and H.~Verlinde,
``Classical and quantum D-branes in 2D string theory,''
arXiv:hep-th/0305194.
%%CITATION = HEP-TH 0305194;%%
}

%\TeschnerQK
\lref\TeschnerQK{ J.~Teschner, ``On boundary perturbations in
Liouville theory and brane dynamics in noncritical string
theories,'' arXiv:hep-th/0308140.
%%CITATION = HEP-TH 0308140;%%
}

\lref\zz{ A.~B.~Zamolodchikov and A.~B.~Zamolodchikov, ``Liouville
field theory on a pseudosphere,'' arXiv:hep-th/0101152.
%%CITATION = HEP-TH 0101152;%%
}

\lref\KazakovCH{ V.~A.~Kazakov and A.~A.~Migdal, ``Recent Progress
In The Theory Of Noncritical Strings,'' Nucl.\ Phys.\ B {\bf 311},
171 (1988).
%%CITATION = NUPHA,B311,171;%%
}

\lref\TakayanagiSM{ T.~Takayanagi and N.~Toumbas, ``A Matrix Model
Dual of Type 0B String Theory in Two Dimensions,''
arXiv:hep-th/0307083.
%%CITATION = HEP-TH 0307083;%%
}
\lref\SW{N. Seiberg and E. Witten, unpublished.}

\lref\GrossAY{
D.~J.~Gross and N.~Miljkovic,
``A Nonperturbative Solution of $D = 1$ String Theory,''
Phys.\ Lett.\ B {\bf 238}, 217 (1990);
%%CITATION = PHLTA,B238,217;%%
%}
%
%\BrezinSS
%\lref\BrezinSS{
E.~Brezin, V.~A.~Kazakov and A.~B.~Zamolodchikov,
``Scaling Violation in a Field Theory of Closed Strings
in One Physical Dimension,''
Nucl.\ Phys.\ B {\bf 338}, 673 (1990);
%%CITATION = NUPHA,B338,673;%%
%}
%
%\GinspargAS
%\lref\GinspargAS{
P.~Ginsparg and J.~Zinn-Justin,
``2-D Gravity + 1-D Matter,''
Phys.\ Lett.\ B {\bf 240}, 333 (1990).
%%CITATION = PHLTA,B240,333;%%
}
\lref\JevickiQN{
A.~Jevicki,
``Developments in 2-d string theory,''
arXiv:hep-th/9309115.
%%CITATION = HEP-TH 9309115;%%
}

\lref\Sennew{ A.~Sen, ``Open-Closed Duality: Lessons from the
Matrix Model,'' arXiv:hep-th/0308068.
%%CITATION = HEP-TH 0308068;%%
}

%\DalleyVR
\lref\DalleyVR{ S.~Dalley, C.~V.~Johnson and T.~Morris,
``Nonperturbative two-dimensional quantum gravity,'' Nucl.\ Phys.\
B {\bf 368}, 655 (1992).
%%CITATION = NUPHA,B368,655;%%
}

%\KutasovSV
\lref\KutasovSV{ D.~Kutasov and N.~Seiberg, ``Number Of Degrees Of
Freedom, Density Of States And Tachyons In String Theory And
Cft,'' Nucl.\ Phys.\ B {\bf 358}, 600 (1991).
%%CITATION = NUPHA,B358,600;%%
}

\lref\GopakumarKI{ R.~Gopakumar and C.~Vafa, ``On the gauge
theory/geometry correspondence,'' Adv.\ Theor.\ Math.\ Phys.\
{\bf 3}, 1415 (1999) [arXiv:hep-th/9811131].
%%CITATION = HEP-TH 9811131;%%
}

\lref\KlebanovHB{ I.~R.~Klebanov and M.~J.~Strassler,
``Supergravity and a confining gauge theory: Duality cascades and
chiSB-resolution of naked singularities,'' JHEP {\bf 0008}, 052
(2000) [arXiv:hep-th/0007191].
%%CITATION = HEP-TH 0007191;%%
}

\lref\MaldacenaYY{ J.~M.~Maldacena and C.~Nunez, ``Towards the
large N limit of pure N = 1 super Yang Mills,'' Phys.\ Rev.\
Lett.\  {\bf 86}, 588 (2001) [arXiv:hep-th/0008001].
%%CITATION = HEP-TH 0008001;%%
}

\lref\VafaWI{ C.~Vafa, ``Superstrings and topological strings at
large N,'' J.\ Math.\ Phys.\  {\bf 42}, 2798 (2001)
[arXiv:hep-th/0008142].
%%CITATION = HEP-TH 0008142;%%
}

\lref\WittenIG{ E.~Witten,
``On The Structure Of The Topological Phase Of Two-Dimensional Gravity,''
Nucl.\ Phys.\ B {\bf 340}, 281 (1990).
%%CITATION = NUPHA,B340,281;%%
}

\lref\MartinecHT{
E.~J.~Martinec, G.~W.~Moore and N.~Seiberg,
``Boundary operators in 2-D gravity,''
Phys.\ Lett.\ B {\bf 263}, 190 (1991).
%%CITATION = PHLTA,B263,190;%%
}

\lref\CachazoJY{ F.~Cachazo, K.~A.~Intriligator and C.~Vafa, ``A
large N duality via a geometric transition,'' Nucl.\ Phys.\ B {\bf
603}, 3 (2001) [arXiv:hep-th/0103067].
%%CITATION = HEP-TH 0103067;%%
}

\lref\BrezinRB{ E.~Brezin and V.~A.~Kazakov, ``Exactly Solvable
Field Theories Of Closed Strings,'' Phys.\ Lett.\ B {\bf 236}, 144
(1990).
%%CITATION = PHLTA,B236,144;%%
}

\lref\DouglasVE{ M.~R.~Douglas and S.~H.~Shenker, ``Strings In
Less Than One-Dimension,'' Nucl.\ Phys.\ B {\bf 335}, 635 (1990).
%%CITATION = NUPHA,B335,635;%%
}

\lref\GrossVS{ D.~J.~Gross and A.~A.~Migdal, ``Nonperturbative
Two-Dimensional Quantum Gravity,'' Phys.\ Rev.\ Lett.\  {\bf 64},
127 (1990).
%%CITATION = PRLTA,64,127;%%
}

\lref\fzz{V.~Fateev, A.~B.~Zamolodchikov and A.~B.~Zamolodchikov,
``Boundary Liouville field theory. I: Boundary state and boundary
two-point function,'' arXiv:hep-th/0001012.
%%CITATION = HEP-TH 0001012;%%
}

\lref\teschner{ J.~Teschner, ``Remarks on Liouville theory with
boundary,'' arXiv:hep-th/0009138.
%%CITATION = HEP-TH 0009138;%%
}

%%%%%%%%%%%%%%%%%%%%%%%%%%%%%%%%%%%%%%%%%%%%%%%%%%%%%%%%%%%%%%%%%%%^M

\Title{\vbox{\baselineskip12pt \hbox{hep-th/0309168}
\hbox{PUPT-2094}  }}
{\vbox{
\centerline{Unitary and Complex
Matrix Models}
\centerline{ as 1-d Type 0 Strings
 } }}
\smallskip
\centerline{I. R. Klebanov,$^{1}$ J. Maldacena,$^2$  and N.
Seiberg$^2$}
\medskip

\smallskip
%\bigskip

\centerline{\it $^1$ Joseph Henry Laboratories, Princeton
University} \centerline{\it Princeton, New Jersey 08544, USA}
\smallskip

\centerline{\it $^2$ Institute for Advanced Study} \centerline{\it
Princeton, New Jersey 08540, USA}
\smallskip
%\bigskip

\bigskip
\noindent We propose that the double scaling behavior of the
unitary matrix models, and that of the complex matrix models, is
related to type 0B and 0A fermionic string theories. The
particular backgrounds involved correspond to $\hat c<1 $ matter
coupled to super-Liouville theory. We examine in detail the $\hat
c=0$ or pure supergravity case, which is related to the double
scaling limit around the Gross-Witten transition, and find
that reversing the sign of the
Liouville superpotential interchanges the 0A and 0B
theories. We also find smooth transitions between
weakly coupled string backgrounds with D-branes, and backgrounds
with Ramond-Ramond fluxes only. Finally, we discuss  matrix models
with multicritical potentials that are conjectured to correspond
to 0A/0B string theories based on $(2, 4k)$ super-minimal models.

\Date{September 2003}

\vfil\eject

\newsec{Introduction}

Recent work on unstable D0-branes of two-dimensional bosonic
string theory
\refs{\mgv\MartinecKA\KlebanovKM\McGreevyEP\SchomerusVV
\AlexandrovNN-\Sennew} has led to reinterpretation of the
well-known large-$N$ matrix quantum mechanics formulation of this
theory (for reviews, see
\refs{\igor\ginsparg\DiFrancescoGinsparg\JevickiQN-\joe}) as exact
open/closed string duality. The open strings live on $N$ unstable
D0-branes; the boundary state of such a D0-brane is a product of
the ZZ boundary state \zz\ for the Liouville field, localized at
large $\phi$, and of the Neumann boundary state for the time
coordinate \refs{\KlebanovKM,\McGreevyEP}. The dynamics of these
open strings is governed by a gauged quantum mechanics of an
$N\times N$ Hermitian matrix $M$ with an asymmetric (e.g. cubic)
potential. This model is exactly solvable since the eigenvalues
act as free fermions. In fact, these eigenvalues {\it are} the
D0-branes. In the double scaling limit \GrossAY, the ground state
of the 2-d bosonic string theory is constructed by filling one
side of the inverted harmonic oscillator potential,
$-\lambda^2/(2\alpha')$, with free fermions up to Fermi level
$-\mu$ as measured from the top of the potential. Since $g_s\sim
1/\mu$, this state has a non-pertubative tunnelling instability.

While this matrix model formulation of the 2-d closed bosonic
string has been known for quite some time \KazakovCH, similar
formulations of NSR strings have been a long-standing problem. In
recent work \refs{\TakayanagiSM,\DouglasUP} a solution of this
problem was found for two-dimensional type 0 strings. Let us
briefly summarize the logic that led to this solution.
Consideration of unstable D0-branes of the type 0B theory indicates
that the dynamics of open strings living on them is again governed
by a gauged Hermitian matrix model, but now with a symmetric
double-well potential. This led the authors of
\refs{\TakayanagiSM,\DouglasUP} to conjecture that the ground
state of the closed 0B string theory corresponds to filling the
potential $-\lambda^2/(4\alpha')$ symmetrically up to Fermi level
$-\mu$. In the continuum formulation the parameter $\mu$ enters
the superpotential of the super-Liouville theory as $\mu e^\phi$.
This explains why in the 0B theory, unlike in the bosonic string,
$\mu$ can have either sign. In fact, this theory has a symmetry
under $\mu \rightarrow -\mu$ \refs{\igor,\joe} which was called
S-duality in \DouglasUP. For either sign of $\mu$ the fermions are
divided symmetrically into two branches in phase space; hence, we
may loosely call this a two-cut eigenvalue distribution.

A matrix model formulation of the 2-d type 0A closed string may be
derived in a similar fashion. The 0A theory has charged D0-branes
and anti D0-branes. The dynamics of open strings on $N+q$
D0-branes and $N$ anti D0-branes is described by $(N+q)\times N$
complex matrix quantum mechanics with $U(N+q)\times U(N)$ gauge
symmetry. Just as the 0B model, this model is exactly solvable in
terms of free fermions and is stable non-perturbatively
\DouglasUP.

In \refs{\TakayanagiSM\DouglasUP
\GaiottoYF\GutperleIJ\KapustinHI\GiveonWN\Harv-\Kitp} these models
were studied further.  In particular, the matrix model
formulations of the 2-d type 0 strings were subjected to a number
of stringent checks vs.\ the continuum worldsheet formulation in
terms of $\hat c=1$ super-conformal field theory coupled to
super-Liouville theory. In this paper we consider further
extensions of these dualities to $\hat c<1$ theories coupled to
super-Liouville. If we turn on relevant operators in the $\hat
c=1$ theory, approporiately dressed by the Liouville field, then
the theory undergoes gravitational RG flow to $\hat c<1$ models
coupled to the super-Liouville theory. Therefore, we expect such
string theories to have matrix model duals closely related to the
ones found for $\hat c=1$. In this paper we indeed argue that type
0B theories are again dual to double-scaling limits of hermitian
matrix models with two-cut eigenvalue distributions (or,
equivalently, of the unitary matrix models), while type 0A
theories are dual to complex matrix models.

For simplicity, we will restrict our discussion to the one-matrix
models. Large $N$ unitary matrix models of this type were solved
in 
\refs{\GrossHE\PeriwalGF\NappiBI\CrnkovicMS\CrnkovicWD\Hollowood-\BrowerMN} 
while the complex matrix models in
\refs{\MorrisBW\DalleyQG\DalleyBR\LafranceWY-\DiFrancescoRU}. The
generic critical behavior of such models is that of the $\hat c=0$
theory (pure supergravity), i.e. of type 0 strings in {\it one}
dimension. The unitary matrix model below the Gross-Witten
phase transition has a two-cut eigenvalue distribution and
above the transition it has one cut.\foot{
Later in the paper we will sometimes
use the phrase ``two-cut hermitian
matrix model'' in referring to the model around this transition, either
above or below the transition where it has one or two cuts.
}
In the double scaling
limit we conjecture that this model is dual to
0B string theory with $\mu >0$ and   $\mu<0$ respectively.
The complex matrix model has a phase transition where the eigenvalues
reach the origin. We conjecture its double scaling limit around this
phase transition to be the dual of 0A string theory.
It was observed long ago that the double scaling limit of the
generic complex
matrix model is equivalent to the one of the generic unitary matrix model
\MorrisBW . Indeed, for $\hat c=0$  we find that a change
in the sign of the left-moving fermion on the wordsheet sends the
0A theory at some value of $\mu$ to the 0B theory at $-\mu$.

 By fine-tuning the potential in the matrix integrals,
we are able to also describe a certain class of
non-unitary $\hat c <0$ SCFT's coupled to super-Liouville theory.
A worldsheet interpretation of these matrix models has been a
longstanding puzzle. In fact, the idea that the two-cut hermitian
matrix models are dual to SCFT's coupled to supergravity was
advanced in the early 90's \refs{\SW,\CrnkovicMS,\CrnkovicWD} but,
as far as we know, was not tested thoroughly.  We will present a
number of consistency checks of this duality conjecture which rely
on the interplay of the 0B and 0A models.

In section 2 we discuss the unitary matrix models,
%with two-cut eigenvalue
%distributions,
reviewing and extending the existing literature on
the subject. Section 3 is devoted to various aspects of the matrix
model resolvent.  It satisfies a loop equation which describes a
Riemann surface.  We demonstrate the discussion with the explicit
solution of the simplest model at tree level. The Riemann surface
also leads to a new insight into the FZZT \refs{\fzz,\teschner} and
the ZZ \zz\ branes in the theory.  In sections 4 and 5 we present
a detailed analysis of the simplest nontrivial theory -- $\hat
c=0$ theory -- pure supergravity, and the first multicritical
point. In section 6 we turn to discussion of the complex matrix model
and its solutions. Section 7 presents a ``spacetime'' picture of
these models.  In section 8-10 we discuss the worldsheet
interpretation of these theories.  We analyze the R-R vertex
operators and uncover interesting dependence on the sign of the
cosmological constant, we discuss the properties of superminimal
models coupled to supergravity and we explore the torus amplitude
in these theories. In section 11 we present our conclusions and
open questions for future research.

Several appendices provide more details for the interested reader.
Some of these details are reviews of known results. In appendix A
we discuss the comparison between the worldsheet and the matrix
model results for the first multicritical point. In appendix B we
mention some properties of superconformal minimal models and a
restriction due to modular invariance on such models. Appendix C
includes an assortment of results about the Zakharov-Shabat
hierarchy of differential operators.  In appendix D and appendix E
we discuss various aspects of the complex matrix model.

\newsec{Unitary Matrix Models }

In this section we study unitary matrix models.
%, which in the
%double-scaling limit are equivalent to hermitian matrix models
%with two-cut eigenvalue distributions. i
The unitary one-matrix
integrals have the form \eqn\unitary{ Z= \int dU \exp \left (
-{N\over \gamma} \Tr V(U+U^\dagger)\right ) \ ,} where $U$ is a
unitary $N\times N$ matrix. The simplest such model, with
potential $\sim \Tr (U+U^\dagger)$, is obtained if one considers
Wilson's lattice action for a single plaquette. This one-plaquette
model was originally solved in the large-$N$ limit by Gross and
Witten \GrossHE . For $\gamma = \infty$ the eigenvalues
$e^{i\theta}$ are uniformly distributed on the circle
parameterized by $\theta$.  As $\gamma$ decreases, the eigenvalue
distribution gets distorted: it starts decreasing in the region
where the potential has a maximum. As $\gamma$ is decreased below
a critical value, $\gamma_c$, a gap appears in the eigenvalue
distribution: this is the third-order large $N$ phase
transition discovered in \GrossHE.
 The generic (and simplest) case
is when the potential has a quadratic maximum. This is labelled as
$k=1$ in the classification of critical points. By further
fine-tuning the potential, Periwal and Shevitz \PeriwalGF\ found
an infinite sequence of multi-critical points labelled by a positive
integer $k$, and found their descriptions in the double-scaling
limit in terms of  the mKdV hierarchy of differential equations.

The double-scaling limit of the simplest unitary matrix model,
$k=1$, with potential $\Tr (U+ U^\dagger)$ is described by the
Painlev\' e II equation \PeriwalGF
\eqn\pain{ 2 f'' -  f^3 +  x f
= 0\ , } where $x\sim (\gamma_c-\gamma) N^{2/3}$. The free energy
$F(x)$ is determined by $F''= f(x)^2/4$. We will identify this
model with pure 2-d supergravity where $x$ is proportional to the
parameter $\mu$ in the super-Liouville interaction, and $-F(x)$ is
the sum over surfaces.

More generally, the $k$-th critical point is described by a
non-linear differential equation for $f(x)$ of order $2k$. The
solution $f$ has the following large $x$ expansion: \eqn\painsol{
f(x)= x^{1/(2k)} \left ( 1 - {2k+1\over 12 k} x^{-(2k+1)/k} +
\ldots \right ) \ , } from which it follows that \eqn\minF{ -
F(x)= - {x^{(2k+1)/k}\over 4 (2 + 1/k) (1+ 1/k)} - {2 k+1\over 24
k} \ln x + \ldots \ . }

In the double scaling limit the physics comes from the coalescence
of two cuts. So, a double scaling limit of a hermitian matrix
model around the point when two cuts meet will lead to the same
free energy as the unitary matrix model. Important further steps
in studying these models were made in 
\refs{\NappiBI,\CrnkovicWD,\Hollowood} where both odd
and even perturbations to the potential were considered. The
complete model is described by two functions, $f(x)$ and $g(x)$,
which in general satisfy coupled equations. An integer $m$
specifies the critical points studied in \CrnkovicWD\ (for even
$m$ the relation to $k$ of \PeriwalGF\ is $m=2k$). In the $m=2$
case the equations are \eqn\coupled{ \eqalign{& 2 f'' + f(g^2-
f^2) + x f = 0\ ,\cr & 2 g'' +  g(g^2- f^2)  +  x g = 0\ ,\cr } }
while the free energy is determined by \eqn\freen{ F''= {1 \over
4} [f(x)^2 - g(x)^2] ~.}

More generally we will need  two functions $F_l$, $G_l$ which are
polynomials in $f$, $g$ and their derivatives. These polynomials
are related to the Zakharov-Shabat hierarchy, which generalizes
the mKdV hierarchy. They are defined by expanding the matrix
resolvent \CrnkovicWD\
\eqn\oper{ O = \langle x| J_3 R | x \rangle
= \langle x| { 1 \over D + Q - \zeta J_3 } | x \rangle =
\sum_{l=-1}^\infty ( -  J_1 F_l - iJ_2 G_l + J_3 H_l ) \zeta^{
-(l+1)} } where $J_i = \sigma_i/2$ with $\sigma_i$ are the
standard Pauli matrices and \eqn\defin{ D = { d \over dx }
~,~~~~~~ Q = \pmatrix{ 0 & f+g \cr f- g & 0} \ .}

Our problem is invariant under a  ``boost'' symmetry of $f$ and $g$.
Therefore, it is natural to define
 \eqn\CDMdef{f\pm g = r e^{\pm \beta }\ ,\qquad\qquad Q = \pmatrix{0 & f+g \cr f- g & 0}  =
 r \pmatrix{ 0 & e^\beta \cr e^{-\beta } & 0 }  \ . }
Similarly, we define a new ``boosted" operator \eqn\newop{ \tilde
O \equiv e^{ - \beta J_3} O e^{\beta J_3} = \sum_{l=-1}^\infty ( -
J_1 R_l - iJ_2 \Theta_l  + J_3 H_l ) \zeta^{ -(l+1)} } where
$R_l$ and $\Theta_l$ are related to $F_l$ and $G_l$ by a boost.
Using
\newop\ and \oper\ we find \eqn\newopr{ \tilde O = { 1 \over D + r
J_1 + ( \omega - \zeta) J_3 } ~,~~~~~~~{\rm with}~~~\omega = \beta'~. }
 We see that undifferentiated
$\beta$ does not appear in $\tilde O$, only $\omega$ and its
derivatives appear. We can see from \newopr\ that a constant shift
in $\omega$ results in a shift in $\zeta $, which in turn
produces a redefinition of the expansion where each $H_l$ gets
mixed with lower $H_l$ terms.

The $2 \times 2$ matrix $O$ obeys the equation \eqn\demand{ [ O ,
D + Q - \zeta J_3 ] =0 } which leads to recursion relations for
$F_l, ~G_l$ and $H_l$ in \oper  . So we can determine them all
from  the lowest ones $G_{-1} =F_{-1}=0$, $H_{-1} =1$ \CrnkovicWD.
Equivalently, we can derive recursion relations for $R_l$,
$\Theta_l$. It turns out that $r \Theta_l$ is always a total
derivative. In fact, $ r \Theta_l = -H'_l$. The recursion
relations then become \eqn\recursionwr{\eqalign{ R_{l+1} & =
\omega R_l - \left( {H'_l \over r} \right)'  + r H_l \ ,\cr H'_{l+1}
& = \omega H'_{l} - r R'_{l}\ . }}
The first few terms are \eqn\firstfew{ \eqalign{& H_{-1} =  1
~,~~~~~~~R_{-1} =0\ ;~ \cr & H_0 = 0 ~,~~~~~~~~~ R_0 = r\ ; \cr &
H_1 =
 - r^2/2~,~~~~~~~~ R_1 = \omega r\ ; ~ \cr
%~,~~~ \Theta_1 = r'
%~,~~~ H_1 = - r^2/2 = - W_1 \cr
& H_2 =  -r^2 \omega~,~~~~~~~ R_2 = -r^3/2 + r \omega^2 + r''\ ; ~
\cr  & H_3 =   {3 \over 8} r^4 - { 3 \over 2} r^2 \omega^2 + { 1
\over 2} r'^2 - r r''\ ,\cr
 & R_3 = - {3 \over 2} r^3 \omega + r \omega^3 + 3 r'
 \omega' + 3 \omega r'' + r \omega''\ . \cr  }}
These equations may also be obtained by ``boosting'' the recursion
relations for $F_l$, $G_l$ in \CrnkovicWD.

It is an interesting exercise to find the terms in $H_l$ and $R_l$
with no derivatives of $r$ or $\omega$.\foot{Since $\omega =
\beta'$ this includes terms with derivatives in terms  of the
original variables.} In terms of the variables \eqn\defrhoangle{
\rho^2 = r^2 + \omega^2 ~,~~~~~~ \cos \varphi = { \omega \over
\rho }~,
% ~~~~~~~ \sin \varphi = {r \over \rho}
} we show in appendix C that
 \eqn\resulteq{\eqalign{ H_l = & - \rho^{l+1} [\cos
\varphi P_l(\cos\varphi) - P_{l+1}(\cos\varphi) ] =- \rho^{l+1} {
\sin^2\varphi \over l+1 }   P'_l(\cos \varphi)\ , \cr R_l = &
\rho^{l+1} \sin \varphi P_l(\cos\varphi) \ .}}

The string equations of the matrix model may be stated as
\CrnkovicWD
 \eqn\strigneq{
 \sum_{l\geq 0} (l+1) t_l G_l =0= \sum_{l\geq 0} (l+1) t_l F_l}
where $t_0 \sim x$. We also assume that $t_l =0$ for $l> m$. These
equations follow from varying the action\foot{This can be proven
as in the KdV case. See  \DiFrancescoGinsparg\ for a nice
discussion of the KdV case. }
 \eqn\action{  S = \int dx  \sum_{l=0}^m t_l H_{l+1} \ .}
 The $l$-th term
correspond to a perturbation in the potential of the schematic
form $V \sim - t_l i^l Tr[ M^{l+2}] $. As pointed out in
\Hollowood, the terms with odd $l$ (and real $t_l$) correspond to
imaginary terms in the potential. Since we are doing an integral,
there is nothing wrong with having these imaginary terms. In fact,
they make the matrix integral more convergent, otherwise an odd
term in the potential would be unbounded below if it ever
dominates.\foot{If we made $t_m$ imaginary for odd $m$ we would
have a real potential. Then the string equation becomes real if we
define $\tilde g \to i g$. But in this case there is no real
solution for $f,~\tilde g$ \Hollowood . }

In all these models the free energy obeys $F'' = -H_1/2 = r^2/4$.
The action \action\ is invariant
under $x$-independent shifts of $\beta$. Therefore, the equation
of motion of $\beta$ is a total derivative and can be integrated
by adding an integration constant $q$. Alternatively, we can view
$\omega = \beta'$ as the independent variable in \action, and add
to it $q \omega$:
 \eqn\qterm{S = \int dx  ( \sum_{l=0}^m t_l H_{l+1}  + q \omega).}
If we assume that this action can be used in a quantum theory such
that $e^{-S}$ is well defined, invariance under $\beta \to \beta +
2\pi i$ leads to the conclusion that $q$ must be quantized. In our
case we are only solving the classical equations coming from this
action, and it is not clear to us which problem the quantum theory
is the answer for. Below we will discuss the physical
interpretation of $q$.

The action \qterm\ leads to the equations of motion
 \eqn\eomaaa{\eqalign{
 &{\delta \over \delta r(x)} \int dx  ( \sum_{l=0}^m t_l H_{l+1}
 + q \omega) = -\sum_{l=0}^m t_l (l+1) R_{l} =0\ ,\cr
 &{\delta \over \delta \omega (x)} \int dx  ( \sum_{l=0}^m t_l
 H_{l+1} + q \omega) = \sum_{l=0}^m t_l (l+1) H_{l} +q =0\ .\cr}}

Suppose we consider the $m$-th model, which has fixed $t_m $. At
first it seems that there are $m$ parameters that we can vary:
$t_l$ with $l=0,\ldots m-1$. However the ability to shift $\omega$
by a constant shows that we can set $t_{m-1}=0$ at the expense of
an analytic change of variables for the rest of the $t_l$; i.e.
the operator that couples to $t_m-1$ is redundant
(this is analogous to a similar operator in
the bosonic
string, which was discussed in \MartinecHT).
 As a result,
there are only $m-1$ operators. If we assign dimension minus one to
$x\sim t_0$, then the operators have dimensions $ 1- l/m$ for $l=
0, \cdots , m-2$. We will later match these operator dimensions
with gravitational dimensions in $(2, 2m)$ super-minimal models
coupled to super-Liouville theory.

If we set $t_l=0$ for all odd $l$ then there exists a solution
with $g=0$. We find that $F_{2k}(f,g=0)$ is the $k$-th member of
the mKdV hierarchy derived for the unitary matrix models in
\PeriwalGF. This corresponds to having an even potential and
considering only even perturbations.  For these even models
$m=2k$. We present examples of models with even $m$ in
sections 4 and 5.

The simplest example of an odd model is the $m=1$ theory.  In this
case \qterm\ becomes
 \eqn\qtermo{S = \int dx  ( -{1\over 2} r^2 \omega - {1\over 2}
 x r^2  + q \omega).}
The equations of motion of this theory are
 \eqn\moneeom{\eqalign{
 &(\omega + x) r=0\cr
 &r^2 = 2q}}
and therefore $u=r^2/4=q/2$.  This ``topological point'' has been
discussed in \CrnkovicWD\ and we now interpret it as associated
with nonzero $q$.  We will say more about the physical
interpretation of $q$ below. Other examples of odd
models are discussed in Appendix C.

\newsec{Comments on the Matrix Model Resolvent}

In this section we will analyze the matrix model in the standard
large $N$ 't Hooft limit, i.e. in the planar approximation. Our main goals
will be to identify the meaning of the parameter $q$ introduced above,
as well as understand the relation between the FZZT
\refs{\fzz,\teschner} and ZZ \zz\ branes in string theory.

Consider the hermitian matrix model with potential $V(M)$.  We
will be interested in two closely related
operators.  The macroscopic loop (the
FZZT brane)
 \eqn\macroloop{W(z)= - {1\over N} \Tr \log (M-z)
 = \lim_{\epsilon \to 0} \left(\int_\epsilon^\infty
 {dl \over l} { 1 \over N} \Tr e^{ l(z -M)} + \log \epsilon\right)}
($\epsilon$ is a UV cutoff) and the resolvent
 \eqn\resolvent{R(z) = {\partial W(z)
 \over \partial z}={1\over N} \Tr {1\over M-z}.}
Without causing confusion we will denote by $W$ and $R$ both the
matrix model operators and their expectation values in the large
$N$ theory.  We will later think of $R(z)dz = dW(z)$ as a
one-form.  It is clear from the expression for $W(z)$ that it can have
an additive ambiguity of ${2\pi i k\over N}$ with integer $k$. If
this is the case, the one-form $R(z)dz$ is not exact.

Using the invariance of the matrix integral under $\delta M =
{1\over M-z}$ we derive the loop equation
 \eqn\loope{ \left\langle \Tr {1\over M-z} \Tr
 {1\over M-z} \right\rangle  +N\left\langle \Tr {V'(M) \over
 M-z}\right\rangle =0 }
The last term can be written as $N V'(z)\left\langle \Tr {1
\over M-z}\right\rangle $ plus a polynomial of degree $n-2$ if
the degree of $V$ is $n$.
 In the large $N$
limit the first term factorizes and we find the loop equation for
the resolvent
 \eqn\reseq{R(z)^2 + V'(z) R(z) - {1\over 4} f(z) = 0,}
where $f(z)$ is a polynomial of degree $n-2$.
 The solution of \reseq\ is
 \eqn\ressol{2 R(z)= - V'(z) \pm \sqrt{V'(z)^2 + f(z)}.}
The cuts in this expression mean that it is a function on a
Riemann surface which is a two-fold cover of the complex plane.
For each value of the parameter $z$ there are two points on the
Riemann surface
 \eqn\riemann{y^2=V'(z)^2 + f(z).}
We will denote them by $P_\pm(z)$.  They differ in the sign of
$y$, $y(P_+(z))=-y(P_-(z))=\sqrt{V'(z)^2 + f(z)}$. The asymptotic
behavior as $z \to \infty$ should be
$R(P_+(z)) \to -{1\over z})$ in the upper sheet. This
determines
  \eqn\ressols{2R(P_{\pm}(z))=  \pm \sqrt{V'(z)^2 +
  f(z)}-V'(z).}

In the planar limit the eigenvalues form cuts. Their density
$\rho(\lambda)= {1\over N} \sum_i \delta(\lambda-\lambda_i)$ is
supported only on the cuts and it can be used to compute
expectation values of operators, e.g.\
 \eqn\Rvev{
  R(z) = {1\over N} \left\langle\Tr
 {1\over M-z}\right\rangle =\int d\lambda {\rho(\lambda) \over
 \lambda - z}\qquad\qquad
 {\rm for ~ } z {\rm ~ not ~ on ~ the ~  cuts } }
from which we find
 \eqn\vevcut{ R(x+ i\epsilon) +
 R(x- i\epsilon)   = 2\CP \int d\lambda
 {\rho(\lambda) \over \lambda - x}\qquad\qquad
 {\rm for ~ } x {\rm  ~ on ~ the ~  cuts }, }
where $\CP\int$ denotes the principal part.

Using the expression for the integral in the planar limit we
define the effective potential of a probe eigenvalue at $z$ away
from the cuts
 \eqn\Veff{V_{eff}(z)=V(z) -2 \int
 d\lambda \rho(\lambda)\log(z-\lambda)= V(z) +  2W(z)  .}
The force on a probe eigenvalue away from the cuts is
 \eqn\Feff{\eqalign{
 F_{eff}(z)&=-V_{eff}'(z)=- V'(z) -  2W'(z)=-V'(z) -2R(z)\cr
 &= -R(P_+(z))+ R(P_-(z)) = - y(z)  .}}
The force on an eigenvalue on the cut, $-{1\over 2}\left(
V'_{eff}(x+ i\epsilon) + V'_{eff}(x- i\epsilon)\right)$ vanishes,
as can be verified using \vevcut\ and \ressols\
 \eqn\forcecut{V'(x)+2\CP \int
 d\lambda {\rho(\lambda) \over \lambda -x} = V'(x)+
 R(x+ i\epsilon) + R(x- i\epsilon) =0. }

We can also analytically continue the FZZT brane \macroloop\ to
the second sheet and distinguish between $W(P_+(z))$ and
$W(P_-(z))$.  Since the analytic continuation of an analytic
function is unique, this generalizes the analytic continuation
discussed in \refs{\TeschnerQK,\MartinecKA}.

This discussion makes it clear that the force $-y(z) dz$ or $R dz$
are one forms on the Riemann surface, while $W$ is the
corresponding potential.  We can calculate the periods of the
force around the $a$-cycles of the Riemann surface
 \eqn\acycle{
{1\over 2\pi i }\oint_{\CC_i} R dz = - {N_i \over N}}
where $\CC_i$ is a contour around the cut $i$ and $N_i$ is the
number of eigenvalues in that cut.  If there is only one cut,
$N_1=N$.  More generally, the residue of $R$ at $z\to \infty$ is
$-1$.  This discussion makes it clear that $W(z)$ of \macroloop\
indeed has additive ambiguities of $2\pi i k/N$ with integer $k$.

Similarly, we can study the periods of $Rdz$ around the $b$-cycles
of the Riemann surface \riemann. Assume for simplicity that there
is only one such cycle; i.e.\ there are only two cuts.  It is a
contour which connects the two cuts in the upper sheet and then
connects them also in the lower sheet.  The period is
 \eqn\bcycle{\hat q = { q \over N} \equiv
{1\over 4\pi i} \oint_{b} y dz ={1\over 2\pi i}
  \oint_{b} R dz .}
This means that $W$ has an additive monodromy of $2\pi i q/N$. We
will later identify it with the flux or the number of D-branes.
This discussion suggest that $q$ must be quantized. It does not
prove it because this discussion applies only in the large $N$
limit with finite $\hat q=q/N$ which is arbitrary.

\subsec{Solution of the simplest model in the planar limit}

Now we will consider the simplest double-well model
with $V'(z)= { 1 \over g}  (z^3-z)$ which leads to the $k=1$
critical behavior.
The second order polynomial $f(z)$ is determined by two
parameters:
\eqn\fform{ f(z) = - {4 \over g} z^2  + f_1 z + f_0 \ ,
}
where the first term is fixed by the condition that $ R(z) \to -1/z$
for large $z$.
 We impose that the sixth order polynomial $y^2(z)$ has
one double zero at $-i\omega$ ($\omega$ can in principle be any complex
number; we will later take it to to be real)
and four simple zeros. This gives  one
relation between the remaing two parameters and leads to the curve
 \eqn\curveI{y^2= {1\over g^2}
( z +i \omega  )^2\,( z^4  - 2i \omega \,z^3 -
 z^2(2 + 3 \omega ^2) + 4 i \,\omega \,z(1+ \omega ^2) +
A  )  \ ,}
with
\eqn\finda{
A =(1 - 4 g + 6 \omega^2 - 5 \omega^4)
}
Turning on the remaining parameter would spread the zero of $y$
into a branch cut.

The scaling limit is obtained when two of the simple zeros of
$y^2$ go to infinity.  Equivalently, we may scale $z$ and $\omega$ to
zero at fixed ratio, while neglecting higher powers of $z$ and
$\omega$.  It is clear from the curve \curveI\ that $A$ should
scale like $z^2$.  Therefore, the critical limit of $g$ is $g=1/4$.
So, we parametrize
$ g = { 1 \over 4} -x/2 $, with $x$ scaling as $\omega$. In this limit the
curve \curveI\ becomes
 \eqn\curveII{
 g_c^2 y^2= -2 ( z + i \omega  )^2\,( (z- i \omega)^2 - 4u )}
with $g_c =1/4$ and
\eqn\equforu{
  4u= x + 2 \omega^2  }
It
is straightforward to compute the period of this $y$ around the
cut between the simple zeros $z_\pm=i \omega \pm 2\sqrt u$
 \eqn\periob{{1 \over 4\pi i} \oint y dz =  8 \sqrt{2}
 \omega u =
 \hat q ={q\over N} .}
It turns out that the second derivative of the free energy
is $\partial_x^2 F = 8 u$.
Up to a rescaling
 of $u,\omega$ and $x$, these are the same
as the equations that follow from \eomaaa\ for the simplest case,
$t_2=1/3$, $t_0 =x/2$ and $q$ nonzero\foot{ The rescaling is
$x_{there} = 2 x_{here}$, $u_{there} = 2 u_{here}$ and $w_{there}
= \sqrt{2} w_{here}$.}. We will discuss these equations in more
detail in  section 4.

Now we solve the two equations \equforu\ \periob .
One can identify $u$ with the second derivative of the free energy.
We will discuss first the case of $q=0$ and then $q\not =0$.
For $\hat q=0$, either $u=0$ or $\omega=0$.
With $\omega=0$ \curveI\ has a double zero at the origin and two
simple zeros at $z_\pm= \pm 2\sqrt u$.  It is natural to run the
cuts in $R$ from these two zeros to infinity.  We take $x=4 u$ to
be positive and identify this with the positive $x$ phase of the
two cut model.  The zero of $y$ at the origin shows that an eigenvalue
at that position feels no force. In string theory
such an eigenvalue corresponds
to an unstable ZZ brane. The potential has a maximum at this point.

With $u=0$ \curveI\ has double zeros at $\pm i \omega$.  Here we
take $\omega$ to be real to describe the negative
$x=- 2\omega^2$ phase of the two/one cut model.  The ends of the
other cut of this model were scaled to infinity and are not
visible in this limit.  The effective potential does not have
stationary points for real $z_0$.  Therefore it is not clear
whether the system has unstable ZZ branes.  However, one can place
some eigenvalues at
 $\pm i  \omega$ where the effective potential is stationary.
We will see that this is equivalent to taking $q \not =0$ and
will be interpreted in the string theory as a charged ZZ brane.
  As we will soon see, branes at
$i\omega$, along the positive imaginary axis, are charged D-branes
and branes at $-i\omega$, along the negative imaginary axis, are
charged anti-D-branes.

Now let us turn to the case that $q \not =0$.
We take $q\to \infty$ with finite $\hat q={q\over N}$ so that its
effects are visible in the planar limit.

Since the period around the $b$-cycle of the Riemann surface is
nonzero, this cycle cannot collapse, and the theory cannot have a
phase transition.

For large positive $x$ we have $u \approx {x \over 4}$ and $\omega
={\hat q \over 4u} \approx  {\hat q \over x}$.  Therefore the
two simple zeros at $z_\pm= i \omega \pm 2\sqrt u\approx i{\hat
q\over x} \pm \sqrt x$ are approximately on the real axis and are
far separated. In this limit it is natural to draw the cuts from
these points to infinity.

As $x$ becomes smaller and negative the two zeros move more into
the complex plane.  They move to the upper (lower) half plane for
$\hat q$ positive (negative).  Let $\hat q$ be positive.  As $x
\to -\infty$ they approach $i \omega \pm 2\sqrt u\approx
i\sqrt{-{x\over 2}} \pm {\hat q\over \sqrt{-2x}}$, i.e.\ they move
up along the positive imaginary axis and approach each other. Here
it is more natural to connect the simple zeros of $y^2$ by a cut
rather than running the cuts to infinity.  This amounts to
performing a modular transformation on the Riemann surface.  In
this configuration the period around this cycle appears to measure
the number of D-branes there.

This gives us a clear picture of why the phase transition at $x=0$
is smoothed out and how the flux (period around the $b$-cycle) is
continuously connected to D-branes (period around the $a$ cycle).

Finally, we would like to suggest that the contour integral \periob\
 of the
force around the $b$ cycle can be interpreted as an imaginary
energy difference between the two Fermi surfaces. In other words,
$l = i q$ is the difference in fermi energies between the two sides.
This puts the
parameters $\hat q$ and $\mu$ on somewhat similar footing and
might suggest a deeper interpretation of these phenomena.

\subsec{ The relation between FZZT and  ZZ branes }

In this subsection, we summarize and slightly extend the
discussion in \refs{\MartinecKA,\TeschnerQK}. As we have argued
above, the operator \macroloop\ should be interpreted as the
insertion of an FZZT brane. It was observed in the CFT analysis
that one can analytically continue the formulae for FZZT branes.
In other words, the branes are labelled by a parameter $s$, and
two different values of $s$ could give the same boundary
cosmological constant $\mu_B$. We will interpret this as the two
sheets of the Riemann surface, in other words we can define the
operators $W(P_+(z))$ and $W(P_-(z))$. It is of interest to find
whether there are points $z_0$ away from the cuts where
$F_{eff}(z_0)=0$. A probe eigenvalue located at these points will
remain stationary, even though it could be unstable. At these
points $y =0$ and $R(P_+(z_0)) = R(P_-(z_0))$. In fact we
 suggest that the operator
 \eqn\ZZo{Z=W(P_+(z_0)) - W(P_-(z_0))}
creates a ZZ brane at $z_0$. In the classical limit of string
theory, the ZZ brane is infinitely heavy and it makes sense to
think of it as an operator, a deformation of the theory.
Similarly, the above analysis in terms of Riemann surfaces is also
valid in the classical limit. Note that since $y$ is zero at $z_0$
we can think of $z_0$ as the position of an infinitesimally short
branch cut. Then \ZZo\ corresponds to integrating $R(z)$ through
the cycle that goes through this infinitesimal cut and the main
cut. In fact, the formula \ZZo\ was inspired by the formulae in
\MartinecKA\
 which express the boundary state of the ZZ brane as a difference
between boundary states of FZZT branes.

Let us consider, for example, the bosonic string. In that case the
FZZT brane parameter $s$ is given by \eqn\values{ \cosh \pi b s =
\mu_B/\sqrt{\mut} } where $\mut$ is, up to an unimportant factor,
the bulk cosmological constant,  and $\mu_B$ is the boundary
cosmological constant. Then the ZZ brane boundary state can be
written as\foot{ We concentrate on the $(1,1)$ ZZ brane.}
\eqn\zzbound{ | D \rangle_{ZZ} = | D (s_+) \rangle_{FZZT} - | D
(s_+) \rangle_{FZZT} ~,~~~~~~~~~s_\pm = i( { 1 \over b} \pm b) }
Note that $\mu_B(s_+) = \mu_B(s_-)$.

One can compute the $\mu$ and $\mu_B$ dependence of the FZZT
branes for the bosonic string with $c=0$. We do this by computing
the disk one point function for the insertion of a bulk
cosmological constant, which equals the derivative of the disk
with respect to the bulk cosmological constant: $\partial_{\mu}
Z_{FZZT} = \langle V_b \rangle $. We compute this using the
formulas in \refs{\fzz,\teschner}. Using \values\ we express the
answer in terms of the boundary cosmological constant to find that
 \eqn\findres{\partial_{\mu_B} D \sim ( 2 \mu_B - \sqrt{\mut})
\sqrt{ \mu_B + \sqrt{\mut} } } which is the expected form of the
singular part of the resolvent for the matrix model corresponding
to bosonic $c=0$. This also gives the force on an eigenvalue. We
see that the force vanishes if $\mu_B^0 = \sqrt{\mut}/2$. One can
check that this is indeed the value of $\mu_B$ that appears for
the FZZT in \zzbound . Furthermore, as $s$ varies continuously
between $ s_+$ and $s_-$, the value of $\mu_B$ goes from $\mu_B^0$
through the cut that starts at $\mu_B = - \sqrt{\mut}$ and back to
$\mu_B^0$.

%We identify the operator \macroloop\ with the $\eta = -1$
%FZZT brane (in the notations of \DouglasUP ) with  $\mu_B=iz$.
%For positive
%$\mu=x$ it exists for real $\mu_B^2 > 2\mu$.
% For negative $\mu=x$ it exists
%for real $\mu_B^2>0$ because there is a singularity at $z=0$. This
%agrees with the expectations based on Liouville theory. In other
%words, for $\mu<0$ the FZZT brane with $\eta = -1$ that we started
%with behaves like the FZZT brane
%with $\eta =1$ in the $\mu>0$ theory.

Now let us consider the superstring theory. There are two
types of FZZT branes distinguished by the sign in the supercharge
boundary condition.
We associate the operator  \macroloop\ to the
 FZZT brane of 0B theory with $\eta = -1$
(in the notations of \DouglasUP ) with  $\mu_B=iz$.
For $\hat c=0$ and $\mu>0$,  one finds  that the
disk expectation value for the $\eta =-1$ brane obeys \Fukuda\
% {\mu_B \over \sqrt{\mut } } = \sinh \pi b s }
% and then for $\hat c =0$ we find $b^{-2} = 2$ and
\eqn\isne{\partial_\mut D_- \sim   \sqrt{ \mut + \mu_B^2 }
 \ .}
This implies
 \eqn\integrexp{\partial_{\mu_B} D_- \sim \mu_B \sqrt{
 \mut + \mu_B^2}}
which is the expected answer; i.e. it is the same as the resolvent
of the two cut model.  On the other hand for the $\eta =+1$ FZZT
brane we find
% \eqn\valmub{ {\mu_B \over \sqrt{\mut } } =
% \cosh \pi b s }
%and we get
\eqn\diskdia{ D_+ \sim \mu_B \mut }
which is analytic in $\mu_B$.

We expect that at negative $\mu$ the expectation value of the
$\eta =-1$ brane will be given by \diskdia .
  This is as expected of the resolvent
of the two cut model for negative $\mu$.

\newsec{The $m=2$ case -- pure supergravity}

In this section we consider in detail the simplest even critical
point of the unitary matrix model, which corresponds to $m=2$ or
$k=1$. As discussed above, we remove $R_1$ away by a shift
of $\omega$, and find the following equations: \eqn\radial{ R_2 +
{1\over 2} x R_0 =0\ , \qquad r\Theta_2 =0\ . } The equation
$r\Theta_2=- H_2'=0$ is a total derivative, which is integrated to
$H_2=-q$. Throughout most of the paper we choose the integration
constant $q$ to be real.  Since $H_2 = - r^2\omega$, $q$ has a
simple interpretation as the boost eigenvalue (the ``rapidity'').
Solving for $\omega$ and inserting it into \radial, we find
 \eqn\stringang{  r'' -{1\over 2}  r^3 + {1\over 2}   x r
 + {q^2 \over r^3} =0\ . }
It follows from the Lagrangian
 \eqn\lagrakoaa{{1\over 2} (r')^2 + {1 \over 8} r^4  - {1\over 4} x
 r^2 + { q^2 \over 2r^2}.}

In \BrowerMN\ the two-cut hermitian matrix model was studied using
different conventions. In \CDMdef\ $g$ was taken to be imaginary;
therefore, $\beta= i\theta$ where $\theta$ is a conventionally
defined angle with periodicity $2\pi$, see \CDMdef . Therefore,
\eqn\stringangnew{ r'' - {r^3\over 2} + {x r\over 2} - r
(\theta')^2 =0\ . } This equation of motion follows from the
Lagrangian \eqn\lagr{ {(r')^2\over 2} + {r^2 (\theta')^2\over 2}+
{r^4\over 8} - {xr^2\over 4} \ . } In Euclidean space the standard
way to impose the constraint on the angular momentum is to add $i
q \theta'$ to the Lagrangian (the factor $i$ is due to the
presence of a single derivative). This method reproduces
\stringang.

Instead, in \BrowerMN\ the ``angular momentum'' $r^2\theta'$ was
taken to be real. This corresponds to imaginary $q$ in \stringang:
$q= il/ 2$, so that
 \eqn\stringangular{ r'' - {1\over 2} r^3 + {1\over 2}x r - {
 l^2 \over 4r^3} =0 \ . }
The real parameter $l$ corresponds to shifting the left ``Fermi
level'' relative to the right one, i.e. to having different
number of eigenvalues in the two wells. This can be seen by
computing the eigenvalue distribution from the leading order
solution (neglecting the $f''$ term). Note that the equation
\stringangular\ comes from the (Euclidean) Lagrangian \lagrakoaa.
We see that with $q=il/2$ this action is not bounded below.  This
implies that we cannot find well behaved non-perturbative
solutions. More precisely, if we look at solutions of $V'(r) =0$,
we see that, for very large $x$, $ r \sim \sqrt{x}$. As we
decrease $x$ we find that at a critical value of $x$ the solution
becomes complex. We can focus on  this critical region by taking
$l\to \infty$ limit and defining the scaled variables
 \eqn\scaledvar{ r =  l^{1/3} +  l^{-1/15} u ~,~~~~~~~~ x ={ 3\over
 2} l^{2/3} + l^{-2/15} y \ .}
Inserting this into \stringangular , taking the $l\to \infty $
limit, and rescaling $y$ and $u \to \tilde u$  by numerical factors,
 we find the Painlev\'e I equation, which is well-known to
describe the double-scaling limit of a one-cut Hermitian matrix model
\refs{\BrezinRB\DouglasVE-\GrossVS}:
 \eqn\painone{
 {1 \over 3} \tilde u'' - \tilde u^2 + y=0 ~,~~~~~~{\rm and}~~~
\partial_y^2 F = { \tilde u \over 2}
\ .}
Thus, the large $l$ limit
corresponds to  removing the eigenvalues from
one side of the potential, and filling the other
side near the top where we recover the single cut critical behavior.

On the other hand, for real $q$ the equation \stringang\ has a smooth
solution. We analyze these solutions below.

\subsec{Solving the ``$q$-deformed'' equation}

Let us examine the solutions of \stringang\ in more detail. For
large positive $x$ we find
 \eqn\stringfice{\eqalign{
 & r(x) =x^{1/2}+ (  4q^2-1)  \left( {1\over 4 x^{5/2}} -
 {-73 + 36q^2 \over 32 x^{11/2}}+ {10657 - 7048\,q^2 +
 1040\,q^4\over 128\, x^{17/2}}   \right.\cr
 &\qquad\qquad \left. -{-13912277 + 10303996\,q^2 - 2156528\,q^4 +
 144704\,q^6\over 2048 x^{23/2}} +\CO( x^{-29/2}) \right) \cr
 &u(x)=r^2(x)/4 ={x\over 4} + \left(q^2  - {1\over 4} \right) \,
 \left[{1\over 2 x^2} +\left( q^2- {9\over 4}  \right) \,\left(-
 {2 \over x^5}+ {14\,\left(q^2  -{21\over 4}  \right) \over
 x^8}\right.\right.\cr
 &\qquad\qquad \left.\left. -{5\,\left( -29 + 4\,q^2 \right) \,
 \left( -83 + 12\,q^2 \right) \over 2\,x^{11}}
 +\CO\left({1\over x^{14}}\right)   \right) \right]\ ,}}
and for large negative $x$,
 \eqn\stringficep{\eqalign{
 &r(x) ={2^{1/4} \sqrt{|q|}\over |x|^{1/4}}\left[ 1- {|q|\sqrt 2
 \over 4 |x|^{3/2}}  + { 5 + 18 q^2\over 32 |x|^{3}}  -
  {|q|\left( 107 + 110 q^2 \right)\sqrt 2 \over 128
  |x|^{9/2}}\right.  \cr
  &\qquad\qquad\qquad\left.
  + {2285 + 13572 q^2 + 6188q^4 \over 2048 |x|^{6}}
  +\CO(|x|^{-15/2})\right] \cr
 &u(x)=r^2(x)/4= {|q |\sqrt 2 \over 4 |x|^{1/2}} -
 {q^2 \over 4\,|x|^2} +
  {5 |q|( 1 + 4\,q^2 )\sqrt 2 \,  \over 64 |x|^{7/2}} -
  {q^2\,\left( 7 + 8\,q^2 \right) \over 8\, |x|^5} \cr
  &\qquad\qquad\qquad\qquad +
  {11\,|q|\,\left( 105 + 664\,q^2 + 336\,q^4 \right)\sqrt 2 \,
    \over 2048 |x|^{13/2}} +  \CO( |x|^{-8})\ .
 }}
The fact that there are terms in the free energy
non-analytic in $q$ at $q=0$
suggests that in the dual string theory there is no R-R vertex operator
that corresponds to turning on $q$ continuously:
had there been a standard vertex operator which couples
to $q$, its $n$-point functions at $q=0$ would have been non-singular.
This suggests that $q$ is quantized.

It is possible to argue that the asymptotic expansion
\stringficep\ matches onto the expansion \stringfice\ as follows.
The differential equation \stringang\ comes from the action
\eqn\action{
S \sim  \int dx \left[ {1 \over 2 } r'^2 +
 { 1 \over 8} (r^2 -  x )^2 +
{ 1 \over 2} { q^2 \over r^2} \right] = \int dx \left [{1 \over 2
} r'^2 + V(r^2)\right ]\ . } This action is bounded below. We can
find an approximate variational solution by neglecting the
derivative term and minimizing the potential for each $x$
independently. This gives a continuous function of $x$. For $q>0$
the function is smooth; for $q=0$ it has a discontinuous first
derivative at $x=0$. Including the second derivative term will
lead to some changes, especially near $x\sim 0$, but a solution
will exist since it is clear that the action
has a minimum.\foot{
It is amusing to also present two exact solutions of our equations
which do not satisfy our boundary conditions. For $q=\pm{1\over
2}$ the problem is solved with $r=\sqrt{x}$, $u=x/4$ and for
$q=\pm{3\over 2}$ it is solved with $r=\sqrt{x+ {4\over x^2} }$,
$u={x\over 4} + {1\over x^2}$.}

Since we started with a well-defined and convergent integral
\unitary, we should end up with a finite and real answer for the
free energy $F$. Therefore, it is natural to expect that the
differential equation has a unique real and smooth solution.
Indeed, the argument above shows this. It is important that this
is the case both for zero and for nonzero $q$. Note that, in order
to select the appropriate solution, it is important to set boundary
conditions both at $x \to \pm \infty $.

In the worldsheet interpretation of these theories we associate
the term of order $|x|^{1-3(h+ b/2)}$ in $u$ with
worldsheets having
$h$ handles and $b$ boundaries.\foot{
Intepretation of this matrix model in terms of world sheets with
boundaries was proposed already in \DalleyBR.}
The $q$
dependence in the expressions \stringficep\stringfice\ is
consistent with this scaling and with a factor of $q^b$ arising
from a surface with $b$ boundaries. This explains why the
coefficient of $x^{1-3n/2}$ is a polynomial in $q$ of degree $n$.

Note also that in terms of the original rank of the matrix, $N$,
before taking the double scaling limit for positive
$x$ we have an expansion with only even powers of $1/N$ while for
negative $x$ we have an expansion with both odd and even
powers.

\subsec{A large $q$ limit}

An interesting limit to consider is $q \to \infty$, $x \to \pm
\infty$ with $t=q^{-2/3} x$ fixed.  This is the 't Hooft limit
with $t^{-3/2}$ being the 't Hooft coupling. After we define
$s=q^{-1/3} r$, equation \stringang\ becomes
 \eqn\stringfica{{2\over q^2} s^3 \partial_t^2 s - s^6 + t s^4 +
 2 =0.}
In the large $q$ limit the first term is negligible, and we end up
with a cubic equation for $v(t)=s^2$
 \eqn\cubicv{ v^3 - t v^2 =2.}
The solution of this equation leads to a free energy $F=q^2 f(t)$
where $v(t)/4=\partial_t^2 f(t)$. This is exactly the expected
behavior in the large $q$ limit when thought of as a large $N$
limit.

For generic $t$ \cubicv\ has three solutions. Only one of them is
real for all $t$ (this is easy to see for $t\approx 0$)
 \eqn\vsol{v(t)={1\over 3}\left[ \,t +
   \left( t^3 + 27 - 3 \sqrt{81 + 6\,t^3}
 \right)^ {\frac{1}{3}}
+  \left( t^3 + 27 +3 \sqrt{81 + 6\,t^3}
 \right)^ {\frac{1}{3}} \right]
 }
(here the branches of the two cube roots should be handled with
care).  For $t> -{3\over  2^{1/3}}$ the arguments of the square
roots are positive and $v(t)$ is real. It is clear from the form
of \vsol\ that the half integer powers of $t$ cancel when we do
the expansion for $t\to \infty $. For $t< - {3\over  2^{1/3}}$ the
arguments of the square roots are negative and the second and
third terms in \vsol\ are complex, but $v(t)$ is real. As we move
in $t$ and the argument of the cubic root moves in the complex
plane and  it is important to keep track of the branch of the cubic
root.  When we expand for very large $t \to - \infty $ we
are are then
saying that $(t^3 + \cdots )^{1/3}  = e^{ \pm i 2 \pi  /3} t +
\cdots $ for the second or third term in \vsol . This  implies
that there is a term where the two square roots add, which leads to
the half integer powers of $t$. The expansions of $v(t)$ for large
negative and large positive $t$
 \eqn\vtexp{v(t)=\cases{{\sqrt {2} \over |t|^{\frac{1}{2}} }
 - \frac{1}{|t|^2} +
  \frac{5 \sqrt {2} } {4 |t|^{7\over 2}} - {4\over |t|^5} +
  \frac{231\sqrt {2} }{32\,
  |t|^{\frac{13}{2}}} + {\CO(\frac{1}{t^8})}\ , & $t<0$\ , \cr
  t+{2\over t^2 } - \frac{8}{{ t  }^5} +\frac{56}{{t  }^8} -
  \frac{480}{{ t  }^{11}} +  \CO(\frac{1}{t^{14}})\ , & $t>0$\ , \cr
 }}
reproduce the highest powers of $q$ in the expansions
\stringficep\stringfice\ after remembering that $u = v/4$.  It is
interesting that one smooth function \vsol\ captures the limiting
behavior of large $q$.

Equation \cubicv\ is obtained from \stringfica\ by neglecting the
derivative term.  In the picture of a particle with coordinate $s$
(or $r$) moving in time $t$ (or $x$) we presented above, this is a
limit where we neglect the acceleration term and keep only the
potential term.  The particle is forced to stay at the stationary
points of the potential.  This suggests that we can take \vsol, or
more precisely $s(t)=\sqrt{v(t)}$, as the starting point of a
systematic expansion of the solution of \stringfica\ in powers of
$1/q^2$, even though usually such an expansion with the highest
derivative term is singular.

We interpret this limit as an 't Hooft limit where only spherical
topologies, perhaps with boundaries, survive.  For $q=0$ the
theory exhibits the Gross-Witten transition \GrossHE\ at $x=0$. Namely
$ F'' = x/4 $ for $x>0$ and $F'' =0$ for $x<0$.  It is known
that this transition can be smoothed by the genus expansion with
$q=0$. Now we see that it can also be smoothed by the expansion in
the 't Hooft parameter for infinite $q$.

An interpretation of this result is the following.  For
negative $t$ we have D-branes and the power of $|t|^{-3/2}$ is the
number of boundaries in the spherical worldsheet.  For positive
$t$ the D-branes are replaced by flux; therefore, we have
spherical worldsheets with insertion of RR fields.  Each RR field
comes with a power of $t^{-3/2}$ but their number must be even,
and hence the expansion in powers of $t^{-3}$.  This is the RR
field  discussed in section 7.
 Note that the power of $t$ agrees with the KPZ
scaling of this operator. It is extremely interesting that the
theory exhibits, both for finite $q$ and infinite $q$, a smooth
transition between the two domains of positive and negative $x$.
This is reminiscent of geometric transitions
\refs{\GopakumarKI\KlebanovHB\MaldacenaYY\VafaWI-\CachazoJY}.
The fact that the theory is smooth at $t=0$, and the strong coupling
singularity is smoothed by the RR flux, may have implications for
QCD-like theories which arise from the conifold with D-branes
\refs{\KlebanovHB\MaldacenaYY\VafaWI-\CachazoJY}.

\newsec{The $m=4$ theory}

Here we briefly discuss the $m=4$ theory which has some new
phenomena compared to the $m=2$ case. In particular, for $x<0$ we
find that the free energy is discontinuous with respect to turning
on the parameter $q$. This happens due to the fact that for
$q=0$ and negative $x$ there exist three different solutions:
the trivial one where
$r=0$, and two non-trivial $Z_2$ symmetry breaking
solutions with the sphere free energy
scaling as $|x|^{5/2}$ (they are related by
the $Z_2$ transformation $\omega\to -\omega$).
For $q=0$ the trivial symmetric solution matches to the positive $x$
solution. For non-vanishing $q$ the positive $x$ solution
matches  one of the non-trivial
symmetry breaking solutions. It is clear that the structure of solutions
gets even richer with increasing $m$.

The basic equations for $m=4$
are \eqn\mfour{ R_4 - {3 x r(x)\over 8} = 0\
,\qquad\qquad H_4 + q=0\ . }
where $R_4,~H_4$ are given in appendix C.
For $x>0$ and $q=0$ there is a
solution with $\omega=0$, and the free energy is given in \minF.
Deformation of this solution by $q$ is straightforward, and we
find \eqn\stringmfpos{\eqalign{
 &r(x) = x^{1/4} +{2\over 3} \left ( {4q^2\over 3}- {5\over
 16}\right )
 x^{-9/4}+ {2\over 3} \left ( -{297\over 128} + {125 q^2\over 9} -
 {608 q^4\over 81} \right ) x^{-19/4}+
 \CO(x^{-29/4})\cr
 &\omega(x) =-{2 q\over 3 x}  + {2q\over 3}\left ({80 q^2\over 27}
 -{5\over 4}\right ) x^{-7/2} +
  \CO(x^{-6})\cr
 &{d^2 F\over dx^2}={r^2(x)\over 4}=
{\sqrt{x}\over 4} + {64 q^2 - 15\over 144} x^{-2}
+ \left ( -{1757\over 2304} +{245 q^2\over 54}
 -{560 q^4\over 243}\right ) x^{-9/2} + \CO(x^{-7})
 }}
 The free energy contains even powers of $q$ only, so we identify
 turning on $q$ in the unitary matrix model
with turning on R-R flux $\sim q$ in the dual string theory.

Now let us consider $x<0$. If $q=0$ then there is an obvious
trivial solution where $r(x)=0$. We have not been able
to find a real deformation of this solution produced by $q$. For
such a deformation one expects $\omega$ to behave as $A|x|^{1/4}$,
while $r\sim |x|^{-3/8}$. However, \mfour\ require that $A^4 < 0$;
therefore, the solution is complex.

There are two other, less obvious, solutions where both $r(x)$ and
$\omega(x)$ are non-vanishing.
One of these solutions has the following asymptotic expansion for
$x\to -\infty$:
\eqn\mfzq{\eqalign{
 &r(x) = (2|x|/7)^{1/4}
- 2^{3/4} \cdot 7^{1\over 4} {5\over 48 |x|^{9/4}}
  - (7/2)^{3/4} {319\over 256 |x|^{19/4}}+ \CO(|x|^{-29/4})\cr
 &\omega(x) =- {\sqrt 3\over 2} (2|x|/7)^{1/4}
- {5 (7/2)^{1/4}\over 32 \sqrt 3 |x|^{9/4}}
- (7/2)^{3/4} {1111\over 1024\sqrt 3 |x|^{19/4}}+ \CO(|x|^{-29/4})\cr
 & {r^2(x)\over 4}= {|x|^{1/2}\over 2\sqrt {14} }
 -{5 \over 48 |x|^2} - (7/2)^{1/2} {2821\over 4608 |x|^{9/2}}
+ \CO(|x|^{-7})
\ , }
}
and the other is related to it by $\omega\to -\omega$.
Since these new solutions break
the $Z_2$ symmetry spontaneously,
in the string interpretation they must involve
background RR fields.  This can be shown explicitly by computing a
nonzero one point function of these RR fields using the matrix model.
The RR fields
which condense in the broken symmetry background are
present in the 0B theory but not
in the 0A theory.\foot{ In the next section we will
discuss a description of the 0A theory by a complex matrix model,
and we will not find any new solutions exhibiting the symmetry breaking
phenomenon.} We also trace the fact that the $m=2$ unitary
matrix model does not exhibit symmetry breaking to the absence
of continuous R-R parameters in this model. As $m$ increases,
so does the number of R-R fields that can condense and produce different
solutions.

Both solutions described above possess real
extensions to non-vanishing $q$. Thus,
we can expand in small $q$ around the
broken symmetry solutions, but not around the symmetric solution.  This
means that it is impossible to vary  $q$ in a continuous fashion around
the symmetric solution. A non-analyticity in $q$
was already obeserved for $m=2$, but for $m=4$ it becomes a
more dramatic phenomenon related to the symmetry breaking.
Just as for $m=2$, in the full nonperturbative $m=4$ theory
there is no vertex operator which changes $q$ in the symmetric background.
This suggests that $q$ is quantized.

The requirement of matching an
$x\to -\infty$ expansion to the solution \stringmfpos\ for large
positive $x$ appears to pick one of the two broken
symmetry solutions for $q>0$, and the other for $q<0$.
 The asymptotic expansion of the $q>0$ solution for
$x\to -\infty$ is
\eqn\stringmf{\eqalign{
 &r(x) = (2|x|/7)^{1/4} + {q\over \sqrt{3} |x|}
- 2^{3/4} \cdot 7^{1\over 4} {15 - 52 q^2\over 144 |x|^{9/4}} + q
\sqrt {42} {165 - 584 q^2\over 864 |x|^{7/2} } +
  \CO(|x|^{-19/4})\cr
 &\omega(x) =- {\sqrt 3\over 2} (2|x|/7)^{1/4} + {2q\over 3 |x|}
- {5 (7/2)^{1/4}\over 32 \sqrt 3} {1+ 32 q^2/3 \over |x|^{9/4}} +
25 q \sqrt {14} {27 - 32 q^2\over 1296 |x|^{7/2} } +
  \CO(|x|^{-19/4})\cr
 & {r^2(x)\over 4}= {|x|^{1/2}\over 2\sqrt {14} }  + {q\over
\sqrt 3 \cdot 7^{1/4} \cdot (2|x|)^{3/4} } +
{ 64q^2 - 15 \over 144 |x|^2}  -
5 (7/2)^{1/4} q {32 q^2-9\over 96\sqrt{3} |x|^{13/4}}
+ \CO( |x|^{-18/4})
 }}
Remarkably, the torus term in the free energy,
${15- 64 q^2\over 144}\ln |x|$,
is the same as for $x > 0$ (see \stringmfpos).
Perhaps this means that the type 0B string theory dual to the $m=4$
model has a symmetry under a change of sign of $x$.

To obtain the solution for $q<0$ we act on the above with the
transformation $q\to -q$, $\omega\to -\omega$. The free energy has
the same structure as in the $k=2$ complex matrix model with
positive $x$ that will be presented in the next section, although
the coefficients are different. The key question is whether the
expansion \stringmf\ for large negative $x$ matches onto the
large positive $x$
expansion in \stringmfpos. We believe that this is the case: a
clear argument in favor of this can be given in the limit of large
$q$.

To take such a limit, we assume $q>0$ and define $\omega = q^{1/5}
h$, $x = q^{4/5} t$, $r^2 = q^{2/5} v$. Then the derivative terms
in \mfour\ are suppressed, and we find \eqn\simple{\eqalign{ & {3
\over 8} v^2 - 3 v h^2 + h^4 - {3 \over 8}  t =0\ , \cr & v h (-
{3 \over 2} v + 2 h^2) = 1\ . }} From the second equation we solve
for $v$ \eqn\solvu{ v_\pm = { 2 \over 3} h^2 \pm \sqrt{ ( {2 \over
3} h^2)^2 -{ 2 \over 3 h} } \ .} From the expansions
\stringmfpos,\stringmf\ we see that $h$ is negative. For $h<0$ we
can only choose the solution $v_+$.

Now we substitute this into the first equation in \simple,
move the square roots to one side of the equal sign, and then
square the equation to remove the square root.
After all this we find the equation
\eqn\eqnforh{
 -12 - 864 h^5  + 448 h^{10}   - 36 h t - 96 h^6  t - 27 h^2  t^2 =0
\ .}
In order to analyze it, we define the
variable
\eqn\newvar{y = h t\ . }
Then the equation becomes quadratic for
$h^5$
\eqn\cuadr{
 -12 - 864 h^5  + 448 h^{10}   - 36 y - 96 h^5 y - 27 y^2 =0
\ .}
Once we find $h$ as a function of $y$ we can find $t = y/h$.
The solutions of this equation are
\eqn\solu{
h^5_\pm = { 54 + 6 y \pm 5 \sqrt{3} \sqrt{28 + 3 (y+2)^2 } \over 56}
\ .}
The solution with
 the plus sign, $h_+^5$,  is always positive and nonzero as a function of $y$.
As we remarked above, we are not interested in solutions with positive $h$.
The solution $h_-$ is non-positive, and it is zero only for
\eqn\zero{
 y_0 = - 2/3
\ .}
As a function of $y$, $h_-$ has a maximum at this point.

As $y\to \pm \infty $ we find that $h_-^5 \sim c_{\pm}
|y|^{1/5}$, where $c_\pm$ is a numerical constant. This implies
that $t$ diverges as $y^{4/5}$ and that $h\sim |t|^{1/4}$. This
agrees with the expansion \stringmf\ which applies for negative
$t$. If we are interested in a solution where $h$ decreases as
$t\to \infty$ it is clear that we should look at the region near
$y_0=-2/3$. This forces us to choose the solution $h_-^5$. As we
change $t$ continuously, we will stay on the $h_-$ branch of \solu\
since the two branches never cross. For $y$ close to $y_0$ we find
that $h \sim y_0/t $, which agrees with the expansion
\stringmfpos\ for large positive $t$.

It is clear that, as $t$ decreases from infinity,
$y$ departs from $y_0$.
%It seems that we could go to the
%left in $y$ or to the right of $y_0$.
Since $h^5_-$ is always negative, in order to have negative values
of $t$ we need to have positive values of $y$. We see then that
the relevant region is $y> y_0$. When $y \to y_0$ we have $t \to
\infty$; when $y=0$ we have $t=0$, and when $y\to +\infty$ we have
$t \to - \infty$. For large  positive $y$, $h_-^5 = -9 y/56 $,
which implies \eqn\implic{ h^4_- = - { 9 \over 56}  t ~,~~~~~~~~~
v = {4 \over 3} h^2 = ( { 2 \over 7} |t|)^{1/2} \ .} This agrees
with the limiting form of the solution for $t\to -\infty$,
\stringmf. Expanding the solution of \solu,\newvar\ further for
large negative $t$, we find \eqn\larq{ h_- = - {\sqrt 3\over 2}
(2|t|/7)^{1/4} + {2\over 3 |t|} - {5 (7/2)^{1/4}\over 3 \sqrt 3
|t|^{9/4}} - {50 \sqrt {14} \over 81 |t|^{7/2} } +
  \CO(|t|^{-19/4})
\ .}
For large positive $t$ we find the expansion
\eqn\larqnew{ h_- =
-{2 \over 3 t}  + { 160\over 81 t^{7/2}} +
  \CO(t^{-6}) \ .
}
The coefficients in these expansions agree with the leading powers for
large $q$ in the expansions of $\omega(x)$ in \stringmf,\stringmfpos.
This shows that an appropriately chosen solution of \simple\
indeed interpolates between the expansion \stringmfpos\ and \stringmf\
in the limit of large $q$. This also suggests that there is a smooth
interpolating solution for finite $q$. Presumably, this could
be checked through numerical work.

\newsec{Complex matrix models}

The complex matrix models are based on an $(N+q)\times N$ complex
matrix $M$ with partition function $Z$, free energy $F$ and
potential $V$
\refs{\MorrisBW\DalleyQG\DalleyBR\LafranceWY-\DiFrancescoRU}
 \eqn\compdef{\eqalign{
 &Z=e^{-F}=\int dM dM^\dagger e^{-{N \over \gamma}
 \Tr V(MM^\dagger)} \ ,\cr
 &V(MM^\dagger) = \sum_{j=1}^p g_j (MM^\dagger)^j \ .}}
Using the $U(N+q)\times U(N)$ symmetry we can bring $M$ to the
form $M_{ij}=\lambda_i \delta_{ij}$ with $\lambda_i \ge 0$. Then
\compdef\ becomes
\eqn\compdefa{\eqalign{
 &Z=e^{-F}=\prod_{i=1}^N \int_0^\infty d\lambda_i
 \lambda_i^{1+2q}
 e^{-{N \over \gamma} V(\lambda_i^2)} \Delta(\lambda^2)^2\cr
 &\qquad\qquad \sim \prod_{i=1}^N \int_0^\infty dy_i y_i^q
 e^{-{N \over \gamma} V(y_i)} \Delta(y)^2\ ,\cr
 &\Delta(y)=\prod_{i>j}(y_i-y_j)\ .}}

Consider first the large $N$ limit with a fixed $\gamma$ -- not
the double scaling limit. The second form of the integral in
\compdefa\ in terms of $y$ is similar to the standard hermitian
matrix model with two exceptions: the factor of $y_i^q$ in the
measure which can be written as a contribution to the potential of
the form $q\sum_i \ln y_i$, and the fact that the domain of $y_i$
is restricted to be positive. Let us now examine the  first form
of the integral in \compdefa .
 We replace $\lambda_i^{1+2q}$ with
$|\lambda_i|^{1+2q}$, and extend the range of the integral over
each $\lambda_i$ to $(-\infty,+\infty)$ (this adds an inessential
constant to $F$ which is independent of the parameters in $V$).
Introducing the eigenvalue density $\rho(\lambda)d\lambda$ the
effective potential is
 \eqn\effpot{\eqalign{
 V_{eff}&=\int_0^\infty d\lambda \rho(\lambda)\left( V(\lambda^2)
 - {(2q +1)\over N} \ln |\lambda|\right) -\int_0^\infty d\lambda d\lambda'
 \rho(\lambda)\rho(\lambda')\ln|\lambda^2-\lambda'^2|\cr
 &= {1\over 2}\int_{-\infty}^\infty d\lambda \tilde \rho(\lambda)
 \left( V(\lambda^2)- {(2q +1) \over N} \ln |\lambda|\right) -{1\over 2}
 \int_{-\infty}^\infty d\lambda d\lambda'
 \tilde \rho(\lambda) \tilde \rho(\lambda')\ln|\lambda-\lambda'|}}
where we have neglected terms of order $1\over N$, we used the
fact that $V$ is invariant under $\lambda \to - \lambda$ and we
defined $\tilde \rho = \rho(\lambda) + \rho(-\lambda)$ with
$\rho(\lambda<0)=0$. In appendix E we derive the loop equation
for this model and we analyze the solution.

For $q \ll N$ \effpot\ is exactly like the standard hermitian
matrix model with an even potential (up to an inessential over all
factor of 2). We conclude that in this case the large $N$ limit of
correlation functions are identical to those of even operators in
the standard hermitian model. Furthermore, if $V(\lambda^2)$ and
$N$ are such that the complex matrix model has one cut that does
not reach $\lambda =0$, then the corresponding hermitian model
would have two cuts. If the cut in the complex matrix model
reaches $\lambda =0$ then the hermitian matrix model has just one
cut. The important lesson is that the planar correlation functions
are the same in the two models. This will be important for us
since we will interpret these two models as the 0A/0B version of
certain string backgrounds.

For $q \sim N$ \effpot\ is like the standard hermitian matrix
model with an additive logarithmic potential.  This is the Penner
model.  We will not explore it further but will assume that $q$ is
finite when $N\to \infty$.

Consider now the $1\over N$ corrections and the double scaling
limit.  We should distinguish three classes of critical behavior
 \item{1.} A cut of eigenvalues ends at generic $\lambda$.
 This critical behavior is described by the standard KdV
 hierarchy of the single cut hermitian matrix model.
 \item{2.} Ends of two different cuts meet at generic $\lambda$.
 This critical behavior is described by the mKdV hierarchy because
 it is the same as in the two cut hermitian matrix model or as in
 the unitary matrix model.
 \item{3.} A cut ends at $\lambda \approx 0$.  In the leading
 order in $1\over N$ (with fixed finite $q $) this critical behavior is
 the same as in two cut models.  However, because of the measure
 factor $|\lambda_i|^{1+2q}$, the $1\over N$ corrections are
 different than in those models.

Since only the third kind of behavior is new, let us focus on it.
The string equation is a differential equation for \eqn\heat{ u=2
\partial_z^2 F(z)} where $z$ is a scaled version of $N$. The
string equation in this case
\refs{\MorrisBW\DalleyQG\DalleyBR-\LafranceWY} is\foot{
For the simplest case,
we provide a derivation of this equation in Appendix D. }
 \eqn\stringq{\eqalign{
 &u\CR^2[u] - {1\over 2} \CR[u] \partial_z^2 \CR[u]
 +{1\over 4} (\partial_z \CR [u])^2 =q^2\cr
 &\CR[u]=\sum_{l=1}^\infty (l +{1\over 2}) t_l Q_l[u] -z}}
where $Q_l[u]$ are the Gel'fand-Dikii differential polynomials
 \eqn\GDpol{\eqalign{
 &Q_0[u]= {1\over 2}\cr
 &Q_1[u] = -{1\over 4} u \cr
 &Q_2[u]= -{1 \over 16} (\partial_z^2 u -3u^2)\cr
 &Q_3[u] = -{1\over 64}(\partial_z^4 u -5 (\partial_z u)^2 -10 u
 \partial_z^2 u + 10 u^3)}}
and $\CR[u]$ is the string equation of the hermitian matrix model
with couplings $t_l$.  Usually the sum over $l$ in $\CR$ is finite
going up to $k$.  Then we normalize the highest $t_l$ such that
the $\CR=u^k + ...$.

For $q=0$ equation \stringq\ is satisfied by a solution of the
differential equation $\CR=0$.  This is exactly the equation of
the standard hermitian one-matrix model in its one cut phase.  The fact that
we find this equation can be easily understood by examining the
expression for the integral \compdefa\ in terms of $y$.  For $q=0$
this is exactly the same integral as in the one-matrix model,
except that the integrals over the eigenvalues are only over
positive $y$.  As long as the support of the eigenvalue
distribution is
away from $y=0$, the effect of the restricted range of integration
is nonperturbative.
Indeed, for $z>0$, it is possible to check that to all orders in the
$1/z$ expansion  \stringq\ leads to
 $\CR=0$.  Clearly, this is not the case
for $z$ negative when the effect of the restricted range of
integration is important. In fact,  the nonperturbative
corrections to the expansion in $1/z$ for $z$ positive are also
sensitive to the bounded range of $y$. Therefore, these
nonperturbative effects are captured by \stringq\ but not by
$\CR=0$.

\subsec{The $k=1$ case}

For a general $q$,
the string equation of the first nontrivial case $k=1$ is
 \eqn\stringfi{\eqalign{
 &u\CR^2[u] - {1\over 2} \CR[u] \partial_z^2 \CR[u]
 +{1\over 4} (\partial_z \CR [u])^2 -q^2=0\cr
 &\CR[u]=u - z.}}
Using the substitution of \MorrisBW,
 \eqn\changef{u(z)=f(z)^2 + z}
the string equation becomes
 \eqn\stringfic{\partial_z^2 f -f^3 -zf +
 {q^2 \over f^3} =0.}
After rescaling $f=2^{-1/6} r$ and $z =2^{-{1/3}}x $ it becomes
the equation we found in the two-cut matrix model, \stringang, but
with the sign of $x$ reversed \MorrisBW. Note also that the
equations \heat,\changef\ for the free energy become, upon the
rescaling, \eqn\heatnew{ {\partial^2 F\over \partial x^2} =
{1\over 4} r^2(x)\ , } up to a non-universal term $\sim x$. This
is consistent with the normalization of the free energy in \freen.

We conclude that the complex matrix model with positive $z$ is the
same as the two cut theory with negative $x$ and vice
versa, i.e. $z=-x$.\foot{The boundary conditions are $u \to z^(1/m) ~(0) $
for $z \to + \infty ~(-\infty)$. We also demand that $\CR $ is positive
and goes to zero as $z \to + \infty$. See appendix D for further
explanation.  }
Therefore, asymptotic expansions of the solution
\stringfice,\stringficep\ apply to the complex matrix model as
well. The complex matrix model with parameter $q$ describes, for
positive $x$, type 0A string theory in the background of $q$
D0-branes. This agrees with the fact that, in \stringficep, the
maximum power of $q$ in each term in the free energy corresponds
to the number of boundaries in a spherical worldsheet. Our
interpretation of the complex matrix model at negative $x$,
described by the expansion \stringfice, is in terms of closed
string in presence of R-R flux.

\def\frac#1#2{{#1\over #2}}

\subsec{The $k=2$ case}

In the $k=2$ case some new phenomena appear compared with $k=1$.
First, let us consider $q=0$. For $z>0$, the $k=1$ complex matrix
model is topological, but the $k=2$ complex matrix model has a
non-trivial genus expansion. Indeed, for $k=2$, the equation is
\DalleyVR
 \eqn\calRe{u \CR^2 - \CR \CR''/2 + (\CR')^2/4 =0\ ,\qquad\quad
 \CR= u^2 - u''/3 -z\ .}
Clearly, it is solved by any solution of $\CR=0$, which for $k=2$
is the Painlev\'e I equation. Around $z\to \infty$ the solution of this
equation has the asymptotic expansion
 \eqn\PLIp{u(z) = \sqrt z -{1\over 24 z^2} -{49\over 1152\
 z^{9/2}} + \CO(z^{-7}),}
but it cannot be continued in a smooth way to negative $z$.

Luckily, there are solutions of \calRe\ which do not satisfy
$\CR=0$ \DalleyVR. Neglecting the derivatives of $u$ we see that
we can solve \calRe\ for negative $z$ without satisfying $\CR=0$
by $u=0$. Expanding around it we find
 \eqn\solxn{u(z)= -  \frac{1}{4\,z^2} -  \frac{225}{32\,z^7}
 - \frac{6906075}{1024\,z^{12}} + \CO(z^{-17}).}
The authors of \DalleyVR\ showed numerically that the two
asymptotic expansions \PLIp\ and \solxn\ are connected through a
real and smooth solution of \calRe.

Generalizing \calRe\ to include $q$, we find \stringq. Now the
asymptotic expansions of the solution are, for positive $z$,
 \eqn\PLIpq{u(z) = \sqrt z +  \frac{|q|}{2\,z^{\frac{3}{4}}}
 -\frac{ 1 + 6\,q^2  }{24\,z^2}  + \CO(z^{-13/4}),}
and for negative $x$,
 \eqn\solxnq{u(z)= (q^2-  \frac{1}{4})\left({1\over z^2}
 +{2(q^2-{9\over 4})(q^2-{25\over 4})\over z^7}
  + \CO(z^{-12})\right).}
The $q\to 0$ limit of this solution is \solxn. Thus, in the
$k=2$ complex matrix
model we do not find a discontinuity at $q=0$ that is present in the
$m=4$ unitary matrix model
even though the two models have the same critical exponents.
The discontinuity of the unitary matrix model
is related to the $Z_2$ symmetry breaking. From the string
point of view, the RR fields that condense in the
0B string are absent from the spectrum of the 0A string
dual to the complex matrix model.\foot{
It is amusing to point out three simple, but unphysical solutions
of \stringq.  For $q=1/2$ it is solved by $u=0$, for $q=3/2$ it is
solved by $u=2/z^2$ and for $q=5/2$ by $u=6/z^2$.  More generally,
it is easy to show that the solution can have double poles with
residues $2$ or $6$.}

We identify this theory with the 0A version of the $(2,8)$
superminimal model for which the KPZ scaling of the $h$ handle and
$b$ boundary surface is $z^{-5(h+b/2-1)/2 -2}$.  The expressions
above are consistent with this identification, with the power of
$q$ being the number of boundaries or insertions of the RR ground
state operator.

As for $k=1$, on one side we have boundaries -- D-branes; on
the other side we have no boundaries but only fluxes, hence
only even powers of $q$ appear.

\newsec{String theory in one dimension}

Here we consider the type 0 theory in one target space dimension.
This theory is also known as pure supergravity or $\hat c=0$
noncritical string theory.  Using standard Liouville conventions
we have $Q={3\over \sqrt 2}$ and $\gamma={1\over \sqrt2}$.
Therefore, the contribution of worldsheets with $h$ handles and
$b$ boundaries scale like $\mu^{3(2-2h-b)/2}$.

The theory has a single closed-string NS operator, the worldsheet
cosmological constant.  In the $(-1,-1)$ picture it is
$e^{\gamma\phi}$.  For $\mu=0$, using free field theory, we find
two candidate R-R operators. In the $(-1/2,-1/2)$ picture they are
given by $V_\pm=\sigma^\pm e^{Q\phi/2}$, where $\sigma^\pm$ are
the two spin fields of the super-Liouville theory. The worldsheet
$(-1)^F$ projection in the 0A theory leaves only $V_+$, and in 0B
it leaves only $V_-$. For nonzero $\mu$ their wave functions are
determined by solving the minisuperspace Schrodinger equations, as
in \DouglasUP. Because of the behavior at $\phi \to \infty$, only
one of the two operators is acceptable. Our conventions are such
that for $\mu<0$ only $V_+$ is acceptable, and for $\mu>0$ only
$V_-$ is acceptable. This means that in the 0A theory there is one
R-R operator $V_+$ for $\mu<0$ and no R-R operator for $\mu>0$. In
the 0B theory the situation is reversed: there is
$V_-$ for $\mu>0$ and no R-R operator for $\mu<0$.

More generally, reversing the sign of $\mu$ can be undone by
transforming the worldsheet Liouville fermions as $\psi \to -\psi$
with $\bar \psi \to \bar \psi$.  Since these are the only fermions
in the problem, this has the effect of changing the sign of the
worldsheet $(-1)^F$ in the R-R sector.\foot{This fact is familiar
from the study of the Ising model.  Changing the sign of the
fermion mass (moving from one side of the transition to the
other), the order and the disorder operators change roles. This
means that we change the sign of the $(-1)^F$ projection in the
R-R sector.} Therefore, it amounts to reversing the projection in
this sector and interchanging 0A and 0B.  We conclude that the 0A
theory with $\mu$ is the same as the 0B theory with $-\mu$. Unlike
the situation in $\hat c=1$ where both theories are invariant
under $\mu \to -\mu$ \DouglasUP, here they are interchanged.

All this can be summarized by a simple ``spacetime''
picture.\foot{We put the word spacetime in quotation marks because we
have only space parametrized by $\phi$.} Recalling the situation
in the $\hat c=1$ theory,
and reducing it to one dimension, we find only one field.
In the 0B theory it is $C$; in the 0A theory it a component of
a gauge field $A_t$ in what was originally the Euclidean time
direction. Since 0A and 0B are related by $\mu \to -\mu$ we can
focus, without loss of generality on 0B\foot{
There is also a RR one form $A_\phi$. }.
 The leading order term
in the action for $C$ is
 \eqn\Clag{\int d\phi \, {1\over 2} e^{2T(\phi) }(\partial_\phi
 C)^2
}
where $ T(\phi) = \mu e^{\gamma \phi} $ is the tachyon field.
The equation of motion is solved by
 \eqn\Ceom{C(\phi)=\int^\phi e^{-2T(\phi')} d\phi'.}
For $\mu >0$ this solution is well behaved as $\phi \to \infty$,
and leads to a ``fluctuating degree of freedom.'' Since it is
linearly divergent as $\phi \to -\infty$, it is not a normalizable
mode and corresponds to a physical vertex operator
\refs{\SeibergEB,\DouglasUP} $V_-$. For $\mu<0$ this solution is
badly behaved for $\phi\to \infty$ and therefore should be
discarded.

We can introduce background charge $q$ at infinity which leads to
background field
 \eqn\backgC{C(\phi)=i q\int^\phi e^{-2T(\phi')} d\phi'.}
This can be done by adding the surface term
 \eqn\Clagsu{i q \int d\phi' \partial_{\phi'} C(\phi')=i
 q\left(C(\infty)-C(-\infty)\right)}
to \Clag. (We view the action \Clag\ as a Euclidean action and
therefore we put an $i$ in front of the topological term
$\partial_\phi C$.  This leads to imaginary classical solution for
$C(\phi)$ as is common in classical solutions of D-instantons.)
For $\mu>0$, integrating out $C(\phi)$ we find ${q^2\over 2} \log
(\Lambda/|\mu|)$ where $\Lambda$ is a cutoff on $\phi$.

For $\mu<0$, the solution of \Ceom\ is badly behaved. In this case
it seems possible to add charged ZZ branes that will absorb the
flux of $C$ and somehow lead to a finite answer. These are charged
$D(-1)$ branes and their number should be precisely equal to the
flux. Therefore, from the string theory point of view it seems
that $q$ should be quantized.

More generally, it is natural to assume that $q$ arises from the
charges which exist in the theory.  These are ZZ branes, which are
localized near $\phi = \infty$.  Therefore, $q$ should be
quantized.  With integer $q$ the surface term \Clagsu\ does not
change if $C(\infty)\to C(\infty)+2\pi$, as expected from the
periodicity of $C$.

$q$ appears as a $\theta$ term, but unlike similar terms, the
theory is not periodic in $q$ because the standard process of pair
creation which can screen its charge does not take place because
the D-branes are locked at infinity.
These D-branes look like instantons.  However, unlike with ordinary
instantons, we do not sum over $q$ but keep it fixed.  There are
two reasons for that.  First,the standard argument
which forces us to sum over $q$ relies on creating an instanton
anti-instanton pair and separating them.  This argument does not
apply here because the D-branes are forced to be at one point in
our ``spacetime.''  Second,
 backgrounds with different values of
$q$ differ by infinite action.

\newsec{R-R physical vertex operators}

In this section we classify the R-R vertex operators, which
distinguish the 0A theory from the 0B theory.

Consider a ``matter'' SCFT with central charge $\hat c$ and an R-R
primary operator with dimension $\Delta>{\hat c \over 16}$. Before
performing any GSO projection there are two such operators
corresponding to the states $|\pm \rangle_M$, where the sign
denotes $(-1)^F$ with $F$ the worldsheet fermion number. It is
convenient to consider the sum and the difference of the zero
modes of the left and right moving matter supercharges,
$G_M={1\over \sqrt 2} (G_{M,Left}+i G_{M,Right})$ and $\bar
G_M={1\over \sqrt 2} (G_{M,Left}-i G_{M,Right})$. They act of these
two states as
 \eqn\mattac{\eqalign{
 & G_M|+\rangle_M= \sqrt{\Delta-{\hat c \over 16}} |-\rangle _M\cr
 & G_M|-\rangle_M=0\cr
 & \bar G_M|-\rangle_M= \sqrt{\Delta-{\hat c \over 16} }|+\rangle
 _M\cr
 &\bar G_M|+\rangle_M =0.}}
There can also be other representations in which
$G_M|+\rangle_M=\bar G_M|-\rangle_M=0$, but $G_M|-\rangle_M$ and
$\bar G_M|-\rangle_M$ do not vanish. Such representations are not
present in the superminimal models. But they do exist in more
generic systems. For example, they exist in $\hat c =1$ and
in  the flat space
critical theory, where we can think of the ``matter part'' as the
superconformal field theory of nine free superfields.  We will
focus on models where all the representations are of the type
\mattac.

We couple this system to Liouville with central charge $\hat
c_L=10-\hat c$.  Physical vertex operators in the R-R sector have
ghosts.  In the $(-{1\over 2} ,- {1\over 2})$ picture the total
dimension of the matter and Liouville is ${5 \over 8}$. Therefore,
the dimension of the Liouville operator $\Delta_L$ needed to dress
the matter operators of dimension $\Delta$ satisfies
$\Delta_L-{\hat c_L \over 16}=-\left(\Delta -{\hat c\over 16}
\right)$.  We denote the two R-R operators with these dimensions
as $|\pm \rangle_L$. (Since $\Delta_L-{\hat c_L \over 16} <0$,
these are operators and not normalizable states \SeibergEB;
nevertheless, we will use the state notation.) The action of the
zero modes of the Liouville supercharges $G_L$ and $\bar G_L$
(which are again linear combinations of left and right moving
supercharges) on them are
 \eqn\liouac{\eqalign{
 &G_L|+\rangle_L= i\sqrt{\Delta-{\hat c \over 16}} |-\rangle _L\cr
 &G_L|-\rangle_L=0\cr
 & \bar G_L|-\rangle_L= i
 \sqrt{\Delta-{\hat c \over 16} }|+\rangle _L\cr
  &\bar G_L|+\rangle_L=0.} }

The total supercharges\foot{ Actually, in order to obey the proper
anticommutation relations we need a cocycle. The proper expression
is $G = G_M \times 1_L + (-1)^{F_M} \times G_L $ where
$(-1)^{F_M}$ is the matter fermion number. We have a similar
expression for $\bar G$. In order not to clutter the equations, we
will suppress the cocycle.} are $G=G_M+G_L$ and $\bar G=\bar
G_M+\bar G_L$ of \mattac\ and \liouac. In the 0A theory the
candidate operators in the $(-{1\over 2} ,- {1\over 2})$ picture
have $(-1)^F=1$, and therefore the allowed operators are $|
+\rangle_M | +\rangle_L$ and $| - \rangle_M | -\rangle_L$.
Imposing that they are annihilated by the total supercharges we
find that there are no such physical states.

In the 0B theory the candidate operators in the $(-{1\over 2} ,-
{1\over 2})$ picture have $(-1)^F=-1$, and therefore the allowed
operators are $| + \rangle_M | -\rangle_L$ and $| - \rangle_M |
+\rangle_L$. Imposing that they are annihilated by $G=G_M+G_L$ and
$\bar G=\bar G_M+\bar G_L$ of \mattac\ and \liouac\ we find one
physical operator
 \eqn\zbvo{| +\rangle_M | -\rangle_L + i | - \rangle_M |
 +\rangle_L.}

As we mentioned above, in theories with  $\hat c >1 $,
there can be matter operators which satisfy \mattac\ with the
interchange of $+\leftrightarrow -$.  For such operators, the
above conclusions about the spectra in 0A and 0B are interchanged.

Consider now the R-R ground states which exists in those SCFT
which do not break supersymmetry.  It has $\Delta={\hat c \over
16}$. Here there is only one state $|+ \rangle _M$ satisfying
 \eqn\mattacz{G|+ \rangle_M= \bar G |+ \rangle _M =0,}
which we take to have even fermion number by convention.  In some
examples like the $\hat c=1$ theory there are two such ground
states $|\pm \rangle_M$ and the discussion below is modified
appropriately.

In the free field description of Liouville there are two states
with $\Delta_L={\hat c_L\over 16}$ which satisfy
 \eqn\liouacz{G_L|\pm \rangle_L= \bar G_L |\pm \rangle _L=0.}
For nonzero $\mu$ only one of them is well behaved as $\phi \to
\infty$ \DouglasUP.  For $\mu<0$ it is $|+ \rangle _L$ and for
$\mu>0$ it is $|- \rangle _L$. Imposing the GSO projection we find
the physical operators.  The 0A theory has an R-R ground state
operator (in the $(-{1\over 2},-{1\over 2})$ picture)
 \eqn\rgphyA{|+\rangle_M|+\rangle_L \qquad {\rm for } \quad \mu<0
 \quad {\rm in} \quad 0A}
and no such operator for $\mu>0$, while the 0B theory has no R-R
ground state operator for $\mu<0$, and
 \eqn\rgphyB{|+\rangle_M|-\rangle_L \qquad {\rm for } \quad
 \mu>0 \quad {\rm in} \quad 0B.}

The spectrum of NS operators in these theories is independent of
the sign of $\mu $, and it is the same in the 0A and 0B theories.
Let us summarize the spectrum of R-R operators.  In the 0A theory
there are no R-R operators for $\mu>0$ and for $\mu<0$ only the
R-R ground state is present.  In the 0B theory with $\mu>0$ there
is a physical vertex operator for each R-R primary.  For $\mu<0$
the spectrum is the same except that the R-R ground state is
absent.

One interesting special case is
 $\hat c=0$.  Here the general picture above is valid,
 but there are no R-R operators other than the ground state.
 Therefore, the spectrum of the 0A theory with one sign of $\mu$
 is the same as the spectrum of the 0B theory with the opposite
 sign of $\mu$.  This is consistent with our general claim that
 $\mu \to -\mu$ exchanges 0A and 0B in this case.

Another interesting case is $\hat c =1$. We can also apply
the above construction, keeping in mind that there will be
two types of matter operators, the ones in \mattac\ and similar
ones with plus and minus interchanged.
 There are two R-R ground states in
 the matter theory with opposite fermion numbers.  For every sign
 of $\mu$ one of them leads to a physical vertex operator in the
 theory coupled to gravity.  This is consistent with the fact that
 these theories are invariant under $\mu \to -\mu$ \DouglasUP.

In nonunitary theories the lowest dimension matter operator is
typically not the identity operator.  Then, a generic perturbation
of the SCFT is given by this operator $x \int d^2\theta
\CO_{\min}$, and upon coupling to Liouville, this operator is
dressed.  We assume that in the conclusions \rgphyA\rgphyB\ about
the existence of the R-R ground state, we simply have to replace
$\mu$ by the coefficient $x$ in this case.

The R-R ground state operators \rgphyA\ and \rgphyB\ represent
fluxes.  If these fluxes can only be induced by D-branes, they are
quantized and cannot be changed in a continuous fashion. Then,
these operators can appear as vertex operators in the perturbative
string theory, and they can appear with quantized coefficients in
the worldsheet theory.  But they cannot exist as standard vertex
operators in the complete nonperturbative theory.

\newsec{Superminimal Models Coupled to Supergravity}

In this section we review some basic facts about
$(A_{p-1}, A_{p'-1})$ superconformal
minimal models and their coupling to super-Liouville theory.

The central charges of the superminimal models are
$$\hat c=1-{2(p-p')^2\over p p'}\ .
$$
The operators are labelled by two positive integers $j$ and $j'$
subject to
 \eqn\mmpc{1\le j'\le p'-1,\qquad 1\le j\le p-1,\qquad jp' \ge j'p
 .}
Their dimensions are
 \eqn\mmpdi{h_{jj'}={(jp'-j'p)^2 -(p-p')^2\over 8 pp'} +
 {1-(-1)^{j-j'}\over 32}.}
Operators with even $j'-j$ are NS operators and those with odd
$j'-j$ are R operators.  The R operator with $jp' = j'p$ is the R
ground state.   It has $h=\hat c/16$.

The super-minimal models are characterized by two positive
integers $p'$ and $p$ subject to
restrictions $p'>p$; $p'-p=0 ~{\rm mod}~ 2$;
if both are odd, they are coprime, and if both are even, then
$p/2$ and $p'/2$ are coprime.  There is also a standard
restriction that if $p$ and $p'$ are even then $(p-p')/2$ is odd
\DiFrancescoXZ\ (see Appendix B). For models with odd $p,p'$ there
is no R state with $h=\hat c/16$; hence these models break
supersymmetry. For models with even $p,p'$ there is such a state,
but we will show that upon coupling to super-Liouville theory, it
does not give rise to a local operator.

When coupled to super-Liouville we need the total $\hat c=10$.
This fixes
 \eqn\Qli{Q=\sqrt{9-\hat c \over 2} = {p+p' \over \sqrt{pp'}}.}
The operator labelled by $(j,j')$ is dressed by $e^{\beta_{j j'}\phi }$
 with exponent
 \eqn\beli{\beta_{jj'}={p+p'-jp'+j'p \over 2\sqrt{pp'}}\ .}
 In the super-Liouville action we may include the coupling
to the lowest dimension operator in the NS sector
of the matter theory, $O_{\min}$,
\eqn\action{
\eqalign{&S(\Phi)={1\over 4\pi}
\int d^2 z d^2 {\theta} \lbrack D_{\theta}\Phi \bar D_{\theta} \bar{\Phi}
     + x \CO_{\min} e^{\beta_{\min} \Phi} \rbrack~, \cr \
&\beta_{\min}={p+p' -2\over 2\sqrt{p p'}}\ ,
\qquad\qquad p, p' ~{\rm even}
\ .\cr}
}
In models with odd $p,p'$ the lowest dimension operator is in the R sector, 
and then we find $\beta_{\min}={p+p' -1\over 2\sqrt{p p'}}$.\foot{
Since models with odd $(p,p')$ break supersymmetry, one might suspect
that their coupling to supergravity leads to theories which are
equivalent to some bosonic minimal models coupled to gravity.
Indeed, there is evidence that
they are equivalent to bosonic $(p,p')$ minimal models
coupled to gravity, with the dressed lowest dimension operator
turned on.
In these bosonic models,
$ Q = {\sqrt 2 (p+p')\over \sqrt{p p'}}$ and $\beta_{\min} =
{p+p'-1\over  \sqrt {2 p p'} }$, so that
${Q\over \beta_{\min}} = {2 (p+p')\over p+p'-1}
$. This is the same as in odd $(p,p')$ superminimal models 
coupled to super-Liouville and perturbed
by the dressed lowest dimension operator from the R sector.
Thus, the scaling of the partition function
is the same. So are the gravitational dimensions of dressed operators.}

Consider now the series of theories $(p=2,p'=2m)$ with
$m=1,2,...$. The standard restriction of \DiFrancescoXZ\ (see
Appendix B) requires that $m$ is even. If we denote $m=2k$, then
the resulting $(2,4k)$ theories, when coupled to super-Liouville,
will match in the 0B case with the critical points of the two-cut
matrix models (these critical points are also labelled by positive
integer $k$ and belong to the mKdV hierarchy found in \PeriwalGF).

Theories with odd $m$ can be obtained by starting with theories
with higher even $m$ and flowing down by turning on a $Z_2$ odd
operator.  In terms of the worldsheet description this is a
superconformal field theory coupled to supergravity which is
perturbed by a R-R operator.  Such theories are not expected to be
massive field theories coupled to Liouville.  The reason is that
the R-R vertex operator involves the ghosts and it is no longer
true that the ghosts and the matter fields are decoupled.  This is
the origin of the difficulty of describing strings in background
R-R fields in the RNS formalism.  However, in the matrix model
there is no such difficulty and the ZS hierarchy allows us to
describe such backgrounds.  In particular, we find new critical
points, the ones with odd $m$, by turning on such R-R operators.
For a discussion of two-cut matrix models with odd
$m>1$, see Appendix C.

The super-Liouville theory is characterized by
 \eqn\Qlik{Q_m = {m+1\over \sqrt{m}}\ .}
The operators are labelled by $(j=1,j'=1,2,...,m-1)$.  The
operators with odd $j'$ are from the NS sector, and those with even
$j'$ are from the R sector.\foot{  We removed from the list the operator
labelled by $j'=m$.  For the SCFT with even $m$ this is the R
ground state which was discussed in the previous section.}  The
Liouville dressings of these operators are determined by
  \eqn\belik{\beta_{j'}={j'+1\over 2\sqrt{m}}.}
The lowest dimension operator is the operator with $j'=m-1$ and
 \eqn\betami{\beta_{\rm min}=\beta_{m-1}={\sqrt{m}\over 2}.}
Since this is the most relevant operator, we will turn it on as
the generic perturbation with coefficient $x$. For even $m$ it is
an NS operator, while for odd $m$ it is an R operator. Therefore,
for even $m$ the theory has $Z_2$ symmetry under which all the R
fields are odd, while for odd $m$ this symmetry is broken.

Standard KPZ/DDK scaling shows that the correlation functions on a
surface with $h$ handles scale like
 \eqn\kpzddk{\langle \prod_i \CO_{j'_i} \rangle_h \sim x^{-{Q_j
(h-1) + \sum_i \beta_{j'_i} \over
 \beta_{\rm min} }}= x^{(-2 -{2\over m})(h-1) - \sum_i {j'_i+1\over
 m}}.}
This matches the discussion of operator dimensions that we discussed
after \eomaaa .
The enumeration of operators, the scaling in \kpzddk\ and the
$Z_2$ assignments agree with that in the double cut (unitary)
matrix model \refs{\PeriwalGF\CrnkovicMS-\CrnkovicWD} with the
string action
 \eqn\ident{S=\int dx \sum_{l=0}^{m} t_l H_{l+1}; \qquad
 t_{m}=1, \qquad t_0\sim x.  }
Here $t_{m-1}$ is redundant and can be shifted away by an
appropriate redefinition of the couplings. As in \MartinecHT, it
can be interpreted as the boundary cosmological constant which is
important only when worldsheet boundaries are present.  More
precisely, it is the boundary cosmological constant in
presence of the FZZT branes \refs{\fzz,\teschner}.
$t_l$ is the coefficient of the operator labelled by
$j'=m-l-1$ in the CFT. For example, $t_{l=0}=x$ is the coefficient
of the lowest dimension operator labelled by $j'=m-1$, $t_{l=m-2}$
is the cosmological constant (the dressed identity operator)
labelled by $j'=1$.

It is simple to repeat this for the 0A theory.  The complex matrix
model is a $Z_2$ orbifold of the two cut model.  It has the same
spectrum of $Z_2$ invariant operators as its parent two cut model
and the $Z_2$ odd operators are absent. The same conclusion about
the spectrum applies to the 0A theory which is an orbifold of the
0B theory.  Therefore, the 0A theory has the same spectrum as the
complex matrix model.

\newsec{The Torus Path Integral}

In this section we compare the matrix model results for
0B and 0A theories with the genus-1 path integral for
supergravity coupled to the
$(A_{p-1}, A_{p'-1})$ superconformal
minimal models.
For even spin structures this was calculated in \BerKlebnew.

As in the bosonic case, we perform the integral over $\phi_0$
first, and it contributes a volume factor $V_L=-(\ln
|x|)/\beta_{\min}$. Integrating over the rest of the modes, we find
\eqn\bas{ {{\cal Z}_{\rm even}^{(S)}\over V_L}=  {1\over 4\pi\sqrt 2}
\int_{\cal F} d^2 \tau  {\tau_2 ^{-3/2}}
   \sum_{(r, s)} |{\cal D}_{r, s}|^{-2}
                            Z_{r, s}^m(\tau, \bar\tau)
} where $Z_{r, s}^m(\tau, \bar\tau)$ is the matter partition
function in the spin structure $(r, s)$. One can represent the
partition functions of the superconformal minimal models as linear
combinations of the partition functions of a compactified scalar
superfield \DiFrancescoXZ.  For the supersymmetric $(A_{p-1},
A_{p'-1})$ series, in each even spin structure there are relations
\DiFrancescoXZ
 \eqn\relation{ \eqalign{&Z_{p,p'}={1\over 2} Z
 ({\cal R}/\sqrt{\ap}=\sqrt{p p'/2}) -
              {1\over 2} Z ({\cal R}/\sqrt{\ap}=\sqrt{p/2p'})
                             \ ,\qquad\qquad p, p' ~{\rm odd}\cr
    & Z_{p,p'}={1\over 2} {\tilde Z} (R/\sqrt{\ap}=\sqrt{p p'/2}) -
                         {1\over 2} Z ({\cal R}/\sqrt{\ap}=\sqrt{p/2 p'})
                           \ ,\qquad\qquad p, p' ~{\rm even}\ ,\cr}
 }
where $\tilde Z$ is the partition function of the ``circle'' $\hat
c=1$ theory, and $Z$ is the partition function of the super-affine
theory. Substituting these expressions into eq. \bas, and
performing the integrals, we find the sum of the contributions of
the $(-, -)$, $(-, +)$ and $(+, -)$ spin structures, each weighted
with the factor $1/2$:\foot{In models with odd $(p,p')$, $x$
is the coefficient of the dressed lowest dimension Ramond operator.
Note that the sum over even spin structures is the same as for bosonic 
$(p,p')$ models coupled to gravity, up to overall coefficient (in the
bosonic case the coefficient is $1/24$ \BerKleb). It is possible that
inclusion of the odd spin structure will give exact agreement
with the bosonic theory. This would provide further evidence that
the odd $(p,p')$ superminimal models coupled to super-Liouville are
equivalent to $(p,p')$ minimal models coupled to Liouville.}
 \eqn\superpart{
 \eqalign{&{\cal Z}_{\rm even} = -{1 \over 16}
 {(p-1)(p'-1)\over (p+p'-1)}\ln| x|\ ,\qquad\qquad p, p' ~{\rm odd}
 \cr &{\cal Z}_{\rm even} = -{1 \over 16}
 {(p-1)(p'-1)+1\over (p+p'-2)}\ln |x|\ ,\qquad\qquad p, p' ~{\rm even}
 \ .\cr }
 }
Note that the answers depend on $\ln |x|$.  Our discussion in
section 8 about the spectrum of these theories shows that the even
spin structures are independent of the sign of $x$.  This sign is
important in the odd spin structure.

The continuum calculation of the odd spin structure is more
difficult and is beyond the scope of this paper. However, even
without calculating it, we can carry out an interesting
consistency check of our 0B and 0A conjectures. Since the odd spin
structure contributes with opposite signs in the 0A and 0B
theories, we can check that the average of their torus partition
sums agrees with \superpart.  Furthermore, since the even spin
structures are independent of the sign of $x$, we should find
 \eqn\conj{{1\over 2} \left ({\cal
 Z}_A (x) + {\cal Z}_B (x)\right ) = {\cal Z}_{\rm even}(|x|)\ .}

Let us consider theories with $p=2, p'= 4k$ where the 1-matrix
models discussed in sections 2-5 and in section 6 describe 0B and
0A theories respectively. In this case the worldsheet computations
lead to
 \eqn\average{ {\cal Z}_{\rm even} = -{1 \over 16} \ln |x| \ , }
independent of $k$.
Let us compare this with the matrix models. We start with the
complex matrix models discussed in section 5, changing the
notation $z\to x$. If we set $q=0$ in \stringq, then the
perturbative solution for $x>0$ is obtained simply by setting
${\cal R}(x)=0$. This is exactly the set of KdV equations that
describe one-cut Hermitian matrix model, except the result is
divided by 2 to eliminate the doubling of the free energy that
appears for symmetric potentials: this is the origin of the factor
$2$ in \heat. It follows that the torus path integral in the
$k$-th multicritical complex matrix model for $x>0$ is the same as
in gravity coupled to the $(2, 2k-1)$ minimal model
which was calculated in \BerKleb. Substituting
$p=2$, $p'=2k-1$ into the result of \BerKleb, we find
 \eqn\answerAp{-{(p-1)(p'-1) \over 24 (p+p'-1)} \ln x
 =-{k-1\over 24k} \ln x.}
Of course, this result can be obtained also by using the KdV
equations ${\cal R}(x)=0$ and integrating \heat.

For $x<0$
we return to \stringq\ and set $q=0$.  It is easy to see that the
asymptotic solution of the equations is $u \approx -{1\over 4
x^2}$ for all $k$ (note the agreement with our $k=1,2$
expressions).  Using \heat\ we learn that
 \eqn\answerA{ {\cal Z}_A= \cases{
 -{k-1\over 24 k} \ln x & $x>0$ \cr
  -{1\over 8} \ln |x| & $x<0$\ . }}
In the  multicritical two-cut models \minF, we find that the torus
path integral for 0B theories is
 \eqn\answerB{ {\cal Z}_B= \cases{
  -{2k+1\over 24 k} \ln x\ & $x>0$\cr
  0& $x<0$.}}
The vanishing result for $x<0$ follows from the fact that,
for $q=0$, the relevant solution for large negative $x$ is the trivial
one, with $r=0$ up to non-perturbative corrections.

Using \answerA,\answerB\
 \eqn\zeven{ {\cal Z}_{\rm even} ={1\over 2} \left ({\cal
 Z}_A (x) + {\cal Z}_B (x)\right )= -{1\over 16} \ln |x| }
in agreement with the worldsheet value \average.  As
expected, it is independent of the sign of $x$.  Using the matrix
model results we get a prediction for the odd spin structure
 \eqn\zodd{ {\cal Z}_{\rm odd} ={1\over 2} \left ({\cal
 Z}_A (x) - {\cal Z}_B (x)\right ) =\cases{
 {k +2 \over 48 k} \ln x & $x>0$ \cr
 &\cr
  -{1\over 16} \ln |x| & $x<0$\ , }}
which does depend on the sign of $x$.  It will be
interesting to check this with an explicit continuum calculation.

\newsec{Conclusions and Future Directions}

In this paper we studied two types of matrix models:
 \item{1.} The unitary matrix model (or, equivalently, the two-cut
 Hermitian matrix model).  In the double scaling limit it is described by the
 differential equations \eomaaa
  \eqn\eomaaaa{\eqalign{
 &{\delta \over \delta r(x)} \int dx  ( \sum_{l=0}^m t_l H_{l+1}
 + q \omega) = -\sum_{l=0}^m t_l (l+1) R_{l} =0\cr
 &{\delta \over \delta \omega (x)} \int dx  ( \sum_{l=0}^m t_l
 H_{l+1} + q \omega) = \sum_{l=0}^m t_l (l+1) H_{l} +q =0.}}
 We identify this model with type 0B string theory in a background
 characterized by $t_l$ and $q$.
 \item{2.} The complex matrix model, which is described in the
 double-scaling limit by the differential equation \stringq
  \eqn\stringqa{\eqalign{
 &u\CR^2[u] - {1\over 2} \CR[u] \partial_z^2 \CR[u]
 +{1\over 4} (\partial_z \CR [u])^2 =q^2\cr
 &\CR[u]=\sum_{l=1}^\infty (l +{1\over 2}) t_l Q_l[u] -z.}}
 We identify this model with type 0A string theory in a background
 characterized by $t_l$ and $q$.

Unlike in the standard Hermitian matrix model, the potentials of
these models are bounded from below, and we expand around their
global minima.  Correspondingly, the differential equations
\eomaaaa\stringqa\ have smooth and real solutions.

We pointed out that, when a superconformal field theory is coupled
to super-Liouville theory to make a superstring theory, there are
in general four independent weak coupling limits.  First, we have
a two-fold ambiguity in the sign of the odd spin structures.  This
leads to the type 0A and type 0B theories.  Second, we can change
the sign of the cosmological constant $\mu \to -\mu$, or more
generally, the sign of the coefficient of the lowest dimension
operator $x \to -x$. In the simplest case of pure supergravity
($\hat c=0$), these theories are identical pairwise: 0A with $\mu$
is equivalent to $0B$ with $-\mu$.  This is not the case, however,
for more general theories.

The recent advances in non-critical string theory have been based
on the idea that the matrix model is a theory of a large number of
ZZ D-branes \refs{\mgv-\Sennew,\TakayanagiSM,\DouglasUP}.  The
identification of the FZZT branes with the integral of the matrix
model resolvent leads to another insight.  The boundary
cosmological constant on this brane can be analytically continued
to take values on a Riemann surface which is a double cover of the
complex plane.  The cuts in the complex plane represent the
eigenvalues; the discontinuity of the resolvent (the derivative of
the FZZT brane with respect to the boundary cosmological constant)
is their density.  This works well for one of the FZZT branes of
the type 0 theory.  It would be nice to identify the other FZZT
brane in the matrix model.

One novelty of the matrix models is that they allow us to analyze
theories with background Ramond-Ramond fields.  This is an
important topic that we have only started to analyze in detail.
Our preliminary investigation has already led to the following
observations:
 \item{1.} Even without turning on the background R-R couplings
 $t_l$, the 0B theory has solutions, e.g.\ \mfzq, which break the
 $Z_2$ symmetry that acts as $-1$ on all the R-R fields.  It
 would be nice to clarify the nature of these solutions in more
 detail\foot{ On the matrix model side, they seem related to
 performing the matrix integral along other contours.}.
 \item{2.} By turning on background R-R fields, the $Z_2$ odd
 $t_l$ couplings, we find new theories.  They involve nontrivial
 couplings of fields and
 ghosts, but their description in the matrix model is as easy as the
 even $t_l$ flows.  Among them we find scaling solutions which are
 described by the odd $m$ equations in the ZS hierarchy.  Unlike
 the even $m$ scaling solutions, these are not superconformal
 matter field theories coupled to supergravity.
 \item{3.}  The parameter $q$ is a particular $Z_2$ odd coupling.
 In the 0B matrix model it appears as an integration constant.  In
 the 0A matrix model it can be introduced by considering a model
 of rectangular matrices.  Alternatively, it can be introduced by
 changing the measure in the complex matrix model of square
 matrices by adding a factor of $(\det MM^\dagger)^{|q|}$ to the
 measure.  In some cases $q$ appears to be related to adding
 D-branes to the system.  This is particularly clear in the
 complex matrix model with rectangular matrices.  From this point
 of view it is natural to assume that $q$ must be quantized.
 However, this conclusion might be misleading.  If $q$ is
 introduced by changing the measure of the 0A model, or if it is
 introduced as an integration constant in the equations of the 0B
 model, we see no reason why it should be quantized.  It would be
 nice to find a clear argument which determines whether $q$ should
 be quantized or not.
 If $q$ is not quantized, we have seen some physical amplitudes
 which are not analytic in $q$ around $q=0$.  Therefore $q$ cannot
 vary in a continuous fashion and its change is not described by
 a standard vertex operator.
 \item{4.}  It is common that for one sign of $x$ the parameter
 $q$ appears with boundaries on the worldsheets and can be
 interpreted as associated with D-branes, while for the opposite
 sign of $x$ only even powers of $q$ appear, and it can be
 interpreted as background R-R flux.  This difference in the behavior
 as $x$ changes sign is consistent with the behavior of the profile
 of the R-R field strength as a function of the Liouville field
 $\phi$.  What is surprising is that the system smoothly
 interpolates between the behavior at positive $x$ and at negative
 $x$.  Such a smooth interpolation between D-branes and fluxes is
 reminiscent of geometric transitions
\refs{\GopakumarKI\KlebanovHB\MaldacenaYY\VafaWI-\CachazoJY}.
 Our solvable
 models provide simple, tractable and explicit examples of this
 phenomenon.
 \item{5.} By examining the planar limit with $x$ positive and $x$
 negative one often finds a phase transition associated with
 nonanalytic behavior of the free energy $F$ \GrossHE.  We have
 seen that the finite $x$ (higher genus) corrections smooth out these
 transitions. Alternatively, we can keep $|x|$ large, i.e.\ continue to
 focus only on spherical worldsheet topologies, but smooth out the
 transition by turning on $q$.  In other words, the transition is
 smoothed out either by including worldsheet handles, or by
 including worldsheet boundaries or nontrivial R-R backgrounds.
 These latter cases provide particularly simple examples of the
 interpolation from D-branes to flux.  They can be seen with
 spherical worldsheets and are described by polynomial equations
 rather than differential equations.  In the 0B theory we have
 interpreted $q$ as a certain period around a cycle in a Riemann
 surface.  The transition associated with the collapse of this
 cycle is prevented by a nonzero period.

We found many situations in which amplitudes vanish without a
simple worldsheet or spacetime explanation.  For example, the
perturbative expansion of $F$ with negative $x$ and $q=0$ vanishes
for some of our solutions of the 0B theory; see e.g.\
\stringficep\ but not \stringmf. Also, the even genus amplitudes
in the expansion \solxnq\ of the 0A theory vanish, but the odd
genus amplitudes are nonzero.  Perhaps these vanishing amplitudes
reflect a deep structure of these theories.

Interesting insights into the bosonic string counterpart of these
systems has been gained by interpreting them as topological
theories \WittenIG. It is possible that a similar topological
structure underlies our examples and our results.  For a possible
starting point for investigating this question, see \CrnkovicWD.

An obvious generalization of the dualities we studied is provided
by multi-matrix model versions of them. It is likely that they can
provide realizations of all $(p,p')$ super-minimal models coupled
to supergravity.  We expect these systems to exhibit a rich
structure which generalizes the phenomena seen in this paper and
in the analysis of the bosonic noncritical string.

We have related  the FZZT brane of the 0B theory with
$\eta =-1$ with the resolvent of the two cut matrix model.
The FZZT brane with $\eta =1$ related to  the resolvent
of the complex matrix model. It would be nice to have a
description of the two branes within the same theory.
It  would also  be nice to have a clearer description of
the relation of the ZZ branes to the FZZT branes.

\bigskip\bigskip\bigskip
\noindent {\bf Acknowledgments:}

We would like to thank E. Martinec and E. Witten for useful
discussions, and C. Johnson for correspondence. 
IRK is grateful to the Institute for Advanced Study
for hospitality during his
work on this paper. The research of JM
and NS is supported in part by DOE grant DE-FG02-90ER40542. The
research of IRK is supported in part by NSF grant PHY-0243680. Any
opinions, findings, and conclusions or recommendations expressed
in this material are those of the authors and do not necessarily
reflect the views of the National Science Foundation.

\appendix{A}{ A simple comparison -- the $m=4$ model}

In this appendix we compute some simple tree level correlation
functions of RR vertex operators using the matrix model results.

The pure supergravity example ($m=2$) discussed above is
particularly simple in that the model has only one coupling
constant which corresponds to turning on the super-Liouville
superpotential $e^\phi$ from the NS sector. As we discussed in the
previous subsection, the R ground state is not present as a
standard deformation of the theory.  Another would be R-operator
is redundant and can be shifted away.

For this reason let us consider the next even critical point,
$m=4$ which we expect to correspond to the $(2,8)$ super-minimal
model coupled to gravity. In this model the effective Lagrangian
is given by \eqn\leff{L = x H_1 + e H_2 + \tilde x H_3 - { 8 \over
15} H_5 \ . } Now there are 2 NS operators, corresponding to
coupling constants $x$ and $\tilde x$, and one RR operator
corresponding to $e$. The coupling $x$ corresponds to the dressed
$j=1, j'=3$ operator of negative dimension, while the coupling
$\tilde x$ to the dressed identity. The coupling $e$ corresponds
to dressed R-R operator $O_R$ with $j=1,j'=2$.  We will set $\tilde
x=0$ and calculate the expansion of the free energy in powers of
$e$.

To perform the sphere calculation we neglect derivatives of $r$
and $\omega$. From the KPZ scaling we expect that  $\langle O_R
O_R\rangle \sim x\ln x$, but we do not find such a term. It follows
that the two-point function vanishes, or is given by a
non-universal term $\sim x$. However, the four point function of
$O_R$ scales as $x^{-1/2}$, in agreement with the KPZ scaling.

\appendix{B}{Super-minimal models with even $p$ and $p'$}

The super-minimal models are characterized by two positive
integers $p'$ and $p$ subject to: $p'>p$, $p'-p=0 ~{\rm mod}~ 2$,
if both are odd, they are coprime, and if both are even, then
$p/2$ and $p'/2$ are coprime.  There is often also a restriction
that if $p$ and $p'$ are even then $(p-p')/2$ is odd.  The purpose
of this appendix is to review parts of the discussion of
\DiFrancescoXZ\ emphasizing why this requirement is needed. Along
the way we will also review some useful facts about the
super-minimal models.

The central charge of the super-minimal model labelled by $(p,p')$
is
 \eqn\centc{\hat c=1-2{(p-p')^2 \over pp'}.}
The operators are labelled by two positive integers $m$ and $m'$
subject to
 \eqn\mmpc{1\le m'\le p'-1,\qquad 1\le m\le p-1,\qquad mp' \ge m'p
 .}
Their dimensions are
 \eqn\mmpdi{h_{mm'}={(mp'-m'p)^2 -(p-p')^2\over 8 pp'} +
 {1-(-1)^{m-m'}\over 32}.}
Operators with even $m'-m$ are NS operators and those with odd
$m'-m$ are R operators.  The R operator with $mp' = m'p$ is the R
ground state.   It has $h=\hat c/16$.

For even $p$ and $p'$ and odd $(p-p')/2$ the R ground state is the
operator labelled by $(m={p\over 2},m'={p'\over 2})$.  For even
$p$ and $p'$ and even $(p-p')/2$ the operator $(m={p\over
2},m'={p'\over 2})$ is in the NS sector and it has
$h=-{(p-p')^2\over 8 pp'} ={\hat c - 1\over 16}$.  This means that
the effective central charge of the theory
\refs{\SeibergEB,\KutasovSV}, which is given in terms of the
lowest dimension operator is $\hat c_{eff}= \hat c - 16
h_{min}=1$, and therefore the density of states of such a theory
is as in the $\hat c=1$ theory.  We conclude that the theory must
have an infinite number of super-Virasoro primaries, and it cannot
be a super-minimal model. Since this argument depends on the
modular invariance of the partition function, let us examine it in
more detail.

The superconformal characters in the different sectors are
\DiFrancescoXZ
 \eqn\char{\eqalign{
 &\chi_\lambda^{NS}(\tau)=\Tr_\lambda q^{L_0-{\hat c \over 16}}=
 \chi^{NS}_{\hat c=1, h=0} (\tau) \left[
 K_\lambda(\tau) - K_{\tilde \lambda} (\tau)\right] \cr
  &\chi_\lambda^{\widetilde {NS}}(\tau) =\Tr_\lambda
  q^{L_0-{\hat c \over 16}}(-1)^F=\chi_\lambda^{NS}(\tau+1)
 =\chi^{\widetilde {NS}}_{\hat c=1, h=0} (\tau) \left[K_\lambda(\tau+1)
  - K_{\tilde \lambda} (\tau+1)\right] \cr
 &\chi_\lambda^{R}(\tau)=\Tr_\lambda q^{L_0-{\hat c \over 16}}=
 \chi^{R}_{\hat c=1, h={1\over 16}} (\tau) \left[
 K_\lambda(\tau) - K_{\tilde \lambda} (\tau)\right] .}}
Here the traces are in the representation labelled by $\lambda=
mp'-m'p$ and we use $\chi^{NS}$ or $\chi^R$ depending on whether
$m'-m $ is even (NS) or odd (R).  In \char\ we use the notation
$\tilde \lambda = mp'+m'p$. $\chi_{\hat c=1,h}$ are the characters
in $\hat c=1$ in the same spin structure and
 \eqn\Kdef{K_\lambda=\sum_{n=-\infty}^\infty q^{(2pp'n +\lambda)^2
 \over 8pp'}}
(we moved a factor of $\eta(\tau)$ from $K$ to the factor that
multiplies it relative to \DiFrancescoXZ).

As a simple consistency checks examine the leading behavior of the
different characters as $q=e^{2\pi i \tau}\to 0$. Since by \mmpc\
$0\le\lambda < pp'$, the leading term in the sum in \Kdef\ is with
$n=0$, and therefore
 \eqn\charl{\eqalign{
 &\chi_\lambda^{NS}(\tau) \to q^{0-{1\over 16} +{\lambda^2 \over
 8pp'}}= q^{h-{\hat c\over 16} } \cr
 &\chi_\lambda^{\widetilde {NS}}(\tau) \to q^{0-{1\over 16}
 +{\lambda^2 \over  8pp'}}= q^{h-{\hat c\over 16} }\cr
 &\chi_\lambda^{R}(\tau) \to  q^{{1\over 16}-{1\over 16}
 +{\lambda^2 \over  8pp'}}= q^{h-{\hat c\over 16} } .}}

Under $\tau \to \tau+1$ each term in the sum \Kdef\ is multiplied
by $e^{\pi i(pp'n^2 + (mp'-m'p)n + {\lambda^2 \over 4pp'})}$.  If
$p$ and $p'$ are even, all the terms have the same phase and $K$
transforms by a phase.  If $p$ and $p'$ are odd, and $m-m'$ is odd
(R representations) again all the terms have the same phase and
$K$ transforms by a phase.  Finally for $p$ and $p'$ odd and
$m-m'$ even (NS representations) the different terms in $K$
transform with the same phase up to $\pm 1$.  Now let us compare
the phase of $K_\lambda$ and $K_{\tilde \lambda}$. Since
${(2pp'n+\lambda)^2-(2pp'n+\tilde \lambda)^2 \over 8pp'} = -n m'p
-{mm' \over 2}$, for R representations (since $mm'$ is even)
$K_\lambda$ and $K_{\tilde \lambda}$ transform with the same
phase.  For NS representations the phase of the terms in
$K_\lambda$ can differ by a minus sign relative to the phase of
the terms in $K_{\tilde \lambda}$.  This is precisely the behavior
expected from the characters because the $L_0$ value of the
different states in the representation differ by integer or half
integer in the NS representations and they differ by an integer in
the R representations.  We conclude that under $\tau \to \tau+1$
the characters transform up to an overall phase as
$\chi_\lambda^{NS}\leftrightarrow \chi_\lambda^{\widetilde {NS}}$
and $\chi_\lambda^R$ are invariant.

Now let us consider the behavior under $\tau \to -{1\over \tau}$.
We use the Poisson resummation formula
 \eqn\poisson{\sum_n e^{-\pi a n^2 + 2\pi i bn}={1\over \sqrt{a}}
 \sum_m e^{-{\pi(m-b)^2\over a}}}
to write
 \eqn\Ktran{K_\lambda(\tau'=-{1\over \tau}) =\sum_n e^{-{2\pi i
 (2pp'n +\lambda)^2 \over 8pp' \tau}}= \sqrt{\tau \over ipp'}
 \sum_n q^{n^2 \over 2 pp'} e^{2\pi i n \lambda \over 2pp'}.
 }

Consider for example the theory with $(p=2,p'=8)$.  It has two NS
representations: the identity with $\lambda=6$, $\tilde \lambda =
10$ and another representation with $\lambda=2$, $\tilde \lambda =
14$.  Using
 \eqn\NSrep{\eqalign{
 \chi_6^{NS}( \tau)=\chi^{NS}_{\hat c=1, h=0} (\tau) \left[
 K_6( \tau) - K_{10} ( \tau)\right] =\chi^{NS}_{\hat c=1, h=0}
 (\tau)\left[\sum_n q^{(16n +3)^2 \over 32} -\sum_n q^{(16n +
 5)^2 \over 32}\right]\cr
  \chi_2^{NS}( \tau)=\chi^{NS}_{\hat c=1, h=0} (\tau) \left[
 K_2( \tau) - K_{14} ( \tau)\right] =\chi^{NS}_{\hat c=1, h=0}
 (\tau)\left[\sum_n q^{(16n +1)^2 \over 32} -\sum_n q^{(16n +
 7)^2 \over 32}\right]\cr
 }}
we easily find
 \eqn\Ktwoeight{\eqalign{
 \chi_6^{NS}(-{1\over \tau})=&\chi^{NS}_{\hat c=1, h=0}
 (-{1\over\tau}) \left[ K_6(-{1\over \tau}) - K_{10}
 (-{1\over \tau})\right]\cr
 =&\sqrt{\tau \over 16 i}\chi^{NS}_{\hat c=1, h=0}
 (-{1\over\tau}) \sum_n q^{n^2 \over 32}  \left[
 e^{2\pi i n 3 \over 16} - e^{2\pi i n 5 \over 16} \right]\cr
  =& \sqrt{\tau \over 4 i}\chi^{NS}_{\hat c=1, h=0}
 (-{1\over\tau}) \sum_k \left[\sqrt{2-\sqrt 2}(q^{(16k+1)^2
 \over 32} -q^{(16k+7)^2 \over  32} )\right.\cr
 &\qquad\qquad\qquad\qquad\qquad \left. - \sqrt{2+\sqrt 2}
 (q^{(16k+3)^2 \over  32} -q^{(16k+5)^2 \over  32} )\right] \cr
  \chi_2^{NS}(-{1\over \tau})=&\chi^{NS}_{\hat c=1, h=0}
 (-{1\over\tau}) \left[ K_2(-{1\over \tau}) - K_{14}
 (-{1\over \tau})\right]\cr
 =&\sqrt{\tau \over 16 i}\chi^{NS}_{\hat c=1, h=0}
 (-{1\over\tau}) \sum_n q^{n^2 \over 32}  \left[
 e^{2\pi i n \over 16} - e^{2\pi i n 7\over 16} \right]\cr
  =& \sqrt{\tau \over 4 i}\chi^{NS}_{\hat c=1, h=0}
 (-{1\over\tau}) \sum_k \left[\sqrt{2+\sqrt 2}(q^{(16k+1)^2
 \over  32} -q^{(16k+7)^2 \over  32} )\right.\cr
 &\qquad\qquad\qquad\qquad\qquad \left. + \sqrt{2-\sqrt 2}
 (q^{(16k+3)^2 \over  32} -q^{(16k+5)^2 \over  32} )\right].
 }}
Therefore, they can be expressed as linear combinations of
$\chi_6^{NS}(\tau) $ and $\chi_2^{NS}(\tau) $.

Let us contrast this with a putative theory with $(p=2,p'=6)$.  It
has two NS representations: the identity with $\lambda=4$, $\tilde
\lambda = 8$ and another representation with $\lambda=0$, $\tilde
\lambda = 12$.  Now
 \eqn\NSrepa{\eqalign{
 \chi_4^{NS}( \tau)=\chi^{NS}_{\hat c=1, h=0} (\tau) \left[
 K_4( \tau) - K_8( \tau)\right] =\chi^{NS}_{\hat c=1, h=0}
 (\tau)\left[\sum_n q^{(6n +1)^2 \over 6} -\sum_n q^{(6n +
 2)^2 \over 6}\right]\cr
  \chi_0^{NS}( \tau)=\chi^{NS}_{\hat c=1, h=0} (\tau) \left[
 K_0( \tau) - K_{12} ( \tau)\right] =\chi^{NS}_{\hat c=1, h=0}
 (\tau)\left[\sum_n q^{(6n )^2 \over 6} -\sum_n q^{(6n +
 3)^2 \over 6}\right]\cr
 }}
and their modular transforms are
 \eqn\Ktwoeighta{\eqalign{
 \chi_4^{NS}(-{1\over \tau})=&\chi^{NS}_{\hat c=1, h=0}
 (-{1\over\tau}) \left[ K_4(-{1\over \tau}) - K_8
 (-{1\over \tau})\right]\cr
 =&\sqrt{\tau \over 12 i}\chi^{NS}_{\hat c=1, h=0}
 (-{1\over\tau}) \sum_n q^{n^2 \over 24}  \left[
 e^{2\pi i n \over 6} - e^{2\pi i n  \over 3} \right]\cr
  =&\sqrt{\tau \over 3 i}\chi^{NS}_{\hat c=1, h=0}
 (-{1\over\tau}) \sum_k\left[q^{(6k+1)^2
 \over 24} -q^{(6k+3)^2 \over 24} \right]\cr
  \chi_0^{NS}(-{1\over \tau})=&\chi^{NS}_{\hat c=1, h=0}
 (-{1\over\tau}) \left[ K_0(-{1\over \tau}) - K_{12}
 (-{1\over \tau})\right]\cr
 =&\sqrt{\tau \over 12 i}\chi^{NS}_{\hat c=1, h=0}
 (-{1\over\tau}) \sum_n q^{n^2 \over 24}  \left[
 1 - (-1)^n \right]\cr
  =& \sqrt{\tau \over 3 i}\chi^{NS}_{\hat c=1, h=0}
 (-{1\over\tau}) \sum_k q^{k^2 \over 6} .
 }}
These are not linear combinations of characters of the theory.
Therefore, the theory with $(p=2,p'=6)$ is not modular invariant.
More generally, when $p$ and $p'$ are even we must assume that
$(p-p')/2$ is odd \DiFrancescoXZ.

\appendix{C}{Some properties of the ZS hierarchy}

In this Appendix we derive the form of the Zakharov-Shabat operators
$R_m$, $H_m$ which is relevant to solving string theory on a sphere.
We also review how the mKdV hierarchy is related to the equation
\stringq\ for the matrix model.

We first recall that,
in terms of $H_m$ and $R_m$, the recursion relations are given by
\recursionwr .
%If we take the $H_{m+1}$ as the Lagrangians, and $r$ and $\w$
%as the fundamental variables then the equations of
%motion of $r$ and $\w$ are proportional to  $R_m$ and
%$W_{m}$ respectively.
It is interesting to solve the recursion relations \recursionwr\
on the sphere, i.e. by dropping the derivative terms and using the
ansatz \eqn\ansatz{ \eqalign{ R_m = & \sum_{l=0}^{[m/2]}  a^m_l
r^{2 l +1 } \w^{ m - 2 l } \cr H_m = & \sum_{l=0}^{ [ (m-1)/2]}
b^m_l r^{ 2l + 2} \w^{ m - 2l -1} }} When we insert this into the
recursion relations \recursionwr\ we can drop the term involving
derivatives of $H$ in the recursion relation for $R_{m+1}$.
Equating coefficients on both sides we find the recursion
relations for the coefficients:
 \eqn\recoeff{\eqalign{ a^{m+1}_l =
& a_{l}^m + b^m_{l-1}\ , \cr b^{m+1}_l = & b_l^m  - { (2 l+1)
\over (2 l+2) } a_l^m \ ,\cr (m-2l) b_l^{m+1} = & -(m - 2 l) a_l^m
+ (m-2l -1)  b_l^m \ .} }
Note that the last two equations have to
be compatible with each other. This implies the equation
\eqn\relationab{
 b_l^m =  - { m- 2 l \over 2 l + 2 } a_l^m
}
This relation can
be stated as \eqn\saysrel{
\partial_r H_m = - \partial_\w R_m
} which is precisely the integrability condition for the second
equation in \recursionwr , when we drop derivative terms in $H_n$ and $R_m$. Note also that once we set $a_0^0 =1$ \recoeff\ implies that $a_0^m=1$.
By demanding that the right hand sides of the first two lines in
\recoeff\ obey \relationab\ we find that \eqn\express{ a^m_l =
(-1)^l { m! \over
 2^{2 l} ( l!)^2 (m-2 l)! }
}
By defining new variables $\rho$ and $\varphi$ through
\eqn\defrhoangle{ \rho^2 = r^2 + \w^2
~,~~~~~~ \cos \varphi = { \w \over \rho }~,~~~~~~~ \sin \varphi =
{r \over \rho}\ , }
we can see from \express\ that $R_m$ and $H_m$ can be written
in terms of Legendre Polynomials
\eqn\resulteq{\eqalign{ H_m = &- \rho^{m+1} [\cos
\varphi P_m(\cos\varphi) - P_{m+1}(\cos\varphi) ] = -
\rho^{m+1} { \sin^2 \varphi \over m+1 }  P'_m(\cos \varphi)\ ,\cr
R_m = & \rho^{m+1} \sin \varphi P_m (\cos\varphi)\ , }}
which is our main result.

This relation can be derived more directly by looking at equation
\oper . In the limit that $r$ and $\w$ are independent of $x$ (i.e.
commute with ${d \over dx}$),
we have the equation \eqn\opera{ { \tilde O} =
\int { dp \over 2 \pi} { 1 \over i p + r J_1 + (\w-\zeta )J_3} =
\int { dp \over 2 \pi } {- ip + \vec v \vec J \over p^2 +
|\vec v|^2/4} =  {\vec v \over | \vec v|} \vec J }
We see that
\eqn\valuesofn{
\eqalign{ & {v_1 \over |\vec v|} =  {
\rho \over |\zeta |} \sin \varphi { 1 \over ( 1 - 2 \cos \varphi
{\rho \over \zeta } + { \rho^2 \over \zeta^2} )^{1/2} } =
{ \zeta \over |\zeta |}
\sum_{l=0}^\infty \zeta^{-l-1} \rho^{l+1} \sin \varphi P_l(\cos
\varphi) \ , \cr
& {v_3 \over |\vec v|}
 = { \rho \over
|\zeta |}  { ( \cos \varphi - { \zeta \over \rho} ) \over ( 1 - 2
\cos \varphi {\rho \over \zeta } + { \rho^2 \over \zeta^2}
)^{1/2} } = { \zeta \over |\zeta|}
\sum_{l=-1}^\infty \zeta^{-l-1} \rho^{l+1} [ \cos
\varphi P_l(\cos \varphi) - P_{l+1}(\cos \varphi) ]\ . \cr }
}
For $\zeta <0$ these equations imply \resulteq .
 Note that in this limit the quantities
$\Theta_l $ in \newop\ are zero since they have
no term without derivatives.

Note that \oper\ implies that
the resolvent, $\Tr[ { 1 \over M - z } ] $,
is proportional to
$ \int^\infty_x dx' Tr[J_3 \tilde O(x')]$ with $z = i \zeta $.
In the planar limit we
can use \valuesofn\ to compute it. We find that the ends of the
cuts are at $ z= i \zeta  = i
\rho e^{ \pm i \varphi} = \pm r + i \omega $.

Let us first consider even $m$ and $q=0$.
The equations on the sphere are
\eqn\sphere{\eqalign{ & \rho \sin \varphi [-\alpha_m x  +
 \rho^{m}
P_m(\cos \varphi)] =0\ , \cr & \rho^{m+1} \sin^2 \varphi P'_m(\cos \varphi)=0
\ , }}
with $ (-1)^{m/2}\alpha_m >0$.
First  note that $\sin \varphi =0$ is
a trivial solution of the equations
with vanishing $r$ and vanishing free energy. Aside from this trivial
solution, the second
equation implies that $P'_{m} (\cos\varphi) =0$.
 For $x>0$ there is
always a solution with $\cos\varphi=0$:
this is the symmetric solution with
vanishing $\omega$.
The number of possible solutions actually grows with $m$.
For $m=2$ the solutions with $\cos\varphi =0,1$ are the only ones.
For $m=4$ and $x<0$ we also find solutions with
$\cos\varphi=\pm \sqrt{3/7}$:
these are the broken symmetry solutions discussed in section 4
(see \stringmf). For $m=6$ and $x<0$ we again find two broken symmetry
solutions with $\cos\varphi = \pm \sqrt{ 15+ 2\sqrt {15} }/\sqrt {33}$;
for $x>0$, in addition to the standard solution with
$\cos \varphi=0$, there are also non-trivial solutions with
$\cos\varphi = \pm \sqrt{ 15- 2\sqrt {15} }/\sqrt {33}$.
In general, the equation $P'_{m} (\cos\varphi) =0$ admits
$m-2$ non-trivial broken symmetry solutions;
some of them are compatible
with $x<0$ and others with $x>0$. For all   the nontrivial
 solutions, $\rho$, $r$,
and $\omega$ scale as $|x|^{1/m}$ for large $|x|$.

If $m$ is odd then let us choose $\alpha_m >0$.
For $x>0$ we
look for solutions of the second equation in \sphere\ with
$P_m(\cos \varphi) >0$ (the solutions with $x\to -x$
are obtained by $\cos\varphi \to - \cos\varphi$).
The number of solutions for a given sign of $x$ is $(m-1)/2$.
After we include derivative terms in the string equations, the
sphere solutions receive higher genus corrections.

Consider, for instance, $m=3$, which is the first non-topological
``odd'' critical point. Here the equations are $R_3 = 5 xr/8$;
$H_3=0$ (see \firstfew). We find the following solution as $x\to
\infty$: \eqn\stringmtpos{\eqalign{
 &r(x) = x^{1/3} -{2\over 9} x^{-7/3} - {1162\over 729} x^{-5}+
 \CO(x^{-23/3})\ ,\cr
 &\omega(x) =-{1\over 2 } x^{1/3} -{2\over 27} x^{-7/3} - {128\over 243} x^{-5}+
 \CO(x^{-23/3})\ ,\cr
 &{d^2 F\over dx^2}={r^2(x)\over 4}={1\over 4} x^{2/3}
 -{1\over 9} x^{-2}- {572\over 729} x^{-14/3}+
  \CO(x^{-22/3})\ .
 }}
Remarkably, for $m$ large enough that there are multiple
solutions, we find that the one loop partition function is
independent of the choice of solution (except for the trivial
solution $r=0$ where it vanishes). The results we find are
consistent with the general formula $F_{\rm torus}= {m+1\over 12
m} \ln x$.

Finally, we extend \firstfew\ by presenting a few
more terms in the Zakharov-Shabat hierarchy, generated via the
recursion relations:
\eqn\fewmore{\eqalign{ R_4 = & \frac{3\,{r}^5}{8} -
3\,{r}^3\,{\w}^2 +
  r\,{\w}^4 - \frac{5\,r\,{r'}^2}{2} +
  12\,\w\,r'\,\w' + 3\,r\,{\w'}^2 -
  \frac{5\,{r}^2\,r''}{2} + \cr &~~~~  6\,{\w}^2\,r'' +
  4\,r\,\w\,\w'' + r^{(4)}
\cr -H_4 = & \frac{-3\,{r}^4\,\w}{2} +
2\,{r}^2\,{\w}^3 -
  2\,\w\,{r'}^2 + 2\,r\,r'\,\w' +
  4\,r\,\w\,r'' + {r}^2\,\w''
\cr R_5 = & \frac{15\,{r}^5\,\w}{8} - 5\,{r}^3\,{\w}^3
+
  r\,{\w}^5 - \frac{25\,r\,\w\,{r'}^2}{2} -
  \frac{25\,{r}^2\,r'\,\w'}{2} +  30\,{\w}^2\,r'\,\w' +
  15\,r\,\w\,{\w'}^2 -\cr &~~~~
   \frac{25\,{r}^2\,\w\,r''}{2} +
  10\,{\w}^3\,r'' -   \frac{5\,{r}^3\,\w''}{2} +
  10\,r\,{\w}^2\,\w'' + 10\,r''\,\w'' +
  10\,\w'\,r^{(3)} +  \cr &~~~~ 5\,r'\,\w^{(3)} +
  5\,\w\,r^{(4)} + r\,\w^{(4)}
\cr - H_5 = & \frac{5\,{r}^6}{16} -
  \frac{15\,{r}^4\,{\w}^2}{4} +
  \frac{5\,{r}^2\,{\w}^4}{2} -
  \frac{5\,{r}^2\,{r'}^2}{4} -
  5\,{\w}^2\,{r'}^2 + 10\,r\,\w\,r'\,\w' +
  \frac{5\,{r}^2\,{\w'}^2}{2} -  \cr &~~~~
  \frac{5\,{r}^3\,r''}{2} +
  10\,r\,{\w}^2\,r'' + \frac{{r''}^2}{2} +
  5\,{r}^2\,\w\,\w'' - r'\,r^{(3)} +
  r\,r^{(4)}
}}

\appendix{D}{ Double scaling limit of the complex matrix model}

In this appendix we derive \stringq\ for the simplest case, $k=1$,
where it reduces to \stringfi. This is a review of the discussion
in \MorrisBW , which is further expanded in \refs{\DalleyQG}.

We start with the complex matrix model \compdef, which reduces to
the integral \compdefa\ \eqn\integral{ Z =\prod_{i=1}^N \int_0^\infty d
y_i y_i^q e^{ - {N \over \gamma} V(y_i)} \Delta(y)^2\ . } We
define orthogonal polynomials with respect to the measure
\eqn\meas{ \int d\mu P_n P_m = \int_0^\infty dy y^q e^{ - {N\over
\gamma} V } P_n P_m = \delta_{n,m} h_n ~,~~~~~P_n = y^n + \cdots}

\subsec{ $q=0$}

Let us first set $q=0$. We derive recursion relations in the usual
way by writing \eqn\recdef{ y P_n = P_{n+1} + s_n P_n + r_n
P_{n-1}, \qquad\qquad r_n= h_n/h_{n-1}\ , } and then writing
 \eqn\recrel{\eqalign{h_{n}^{-1}\langle n |
 V' |n \rangle &=  { \gamma \over N} {P_n(0)^2 \over h_n}
 e^{ - { N \over \gamma} V(0) } = \Omega_n\ ,
\cr h_{n-1}^{-1}  \langle n -1 | V' | n \rangle & = { \gamma n
\over N} + { \gamma \over N} {P_{n-1}(0) P_n(0) \over h_{n-1} }
e^{ - { N \over \gamma} V(0) }
 \equiv x_n+ \tilde \Omega_n\ ,
 }}
where $\langle m|n\rangle = h_n \delta_{mn}$, and $x_n = { \gamma
n \over N} $. $\Omega$ and $\tilde \Omega$ are defined in terms of
the values of the polynomials at zero
 \eqn\omomt{ \Omega_n \equiv{ \gamma \over N} {P_n(0)^2 \over h_n}
 e^{ - { N \over \gamma} V(0) } , \qquad \qquad \tilde \Omega_n \equiv
 { \g \over N} {P_{n-1}(0) P_n(0) \over h_{n-1} }  e^{
 - { N \over \g} V(0) }.
 }
They satisfy
 \eqn\eqnsforomega{\eqalign{& \tilde \Omega_n + \tilde
\Omega_{n+1} = - s_n \Omega_n \ ,\cr & r_n \Omega_n \Omega_{n-1} =
\tilde \Omega^2_n \ .}}

The simplest potential, which arises for the $k=1$ model, is $V(y)
=- y + y^2/2$ and then we find that \recrel\ are
 \eqn\recrelsim{ -1+ s_n = \Omega_n ~,~~~~~~ r_n - x_n  = \tilde
 \Omega_n \ .}
We can eliminate $s_n$ from this equation and substitute in
\eqnsforomega\ to end up with
 \eqn\recfi{\eqalign{
 &r_n - x_n  = \tilde \Omega_n \ ,\cr
 & \tilde \Omega_n + \tilde \Omega_{n+1} = - (\Omega_n+1)\Omega_n \ ,\cr
 & r_n\Omega_n \Omega_{n-1} = \tilde \Omega^2_n\ .}}
When the cut is far from $y =0$, the polynomials are small at the
origin; therefore, we find $\Omega  = \tilde  \Omega_n =0$ and
recover the standard hermitian matrix model equations.

One can then combine these equations with \recrel\ to obtain the
string equations.  In the simplest model this works as follows.
Let us start by considering the planar limit.  The equations
\eqnsforomega\ and \recrelsim\ (or equivalently \recfi) have three
solutions on the sphere:
 \item{1.} $\Omega=\tilde \Omega =0$, $s=1$, $r=x$.
 \item{2.} $\Omega={1\over 3} (-2 + {\sqrt{1 + 12\,x}})$,
 $\tilde \Omega={1\over 18}( 1 - 12\,x + {\sqrt{1 + 12\,x}})$,
 $s={1\over 3} (1 + \sqrt{1 + 12\,x})$ and $r = {s^2\over 4} ={1\over 18}
 (1 + 6\ x +\sqrt{1 + 12\ x})$.
 \item{3.} $\Omega={1\over 3} (-2 - {\sqrt{1 + 12\,x}})$,
 $\tilde \Omega={1\over 18}( 1 - 12\,x - {\sqrt{1 + 12\,x}})$,
 $s={1\over 3} (1 - \sqrt{1 + 12\,x})$ and $r = {s^2\over 4} ={1\over 18}
 (1 + 6\ x -\sqrt{1 + 12\ x})$.
At this point we need more physical input.  It arises from \omomt\
and the interpretation of $\Omega$ in terms of $P_n(0)$.  First,
it is clear that $\Omega \ge 0$.  This rules out the third
solution and allows the second solution only for $x\ge {1\over
4}$. Second, it is clear that we cannot have $P_n(0)=0$ for all
$n$.  Therefore we cannot take the first solution for all $x$.  We
conclude that for $0<x\le {1\over 4}$ we should take the first
solution and for ${1\over 4}\le x \le 1 $ the second solution.  In
terms of the eigenvalues, for $x<{1\over 4} $ they stay away from
the origin so that $\Omega=0$; but for ${1\over 4}\le x \le 1 $
they reach the origin so that $\Omega\not=0$.

Now, consider the double scaling limit around the transition point
$x=1/4$,
 \eqn\scal{ r = {1\over 4} -  \epsilon \hat u, ~~~~~~ x = {1\over 4} - \epsilon
 \tilde z
 ,~~~~~~ \Omega = \epsilon \hat \Omega , ~~~~~~
 \tilde \Omega = - \epsilon \tilde {\cal R}, ~~~~~~ N=\g \epsilon^{-3/2}.}
Equations \recfi\ become
 \eqn\equa{\eqalign{
 & \tilde \CR = \hat u- \tilde z \ ,\cr
 &2 \tilde \CR - \epsilon^{1/2} \tilde \CR'
  + { \epsilon \over 2} \tilde \CR'' +\CO(\epsilon^{3/2}) = \hat \Omega + \epsilon \hat
 \Omega^2 \ ,\cr
 &({1\over 4} - \epsilon \hat u ) \hat \Omega (\hat \Omega +
 \epsilon^{1/2} \hat \Omega' + { \epsilon \over 2} \hat \Omega'')
 = \tilde \CR ^2  +\CO(\epsilon^{3/2})\ ,\cr
 }}
where derivatives are evaluated with respect to $\tilde z$.
Solving the second equation for $\hat \Omega$, and substituting
into the last one, we find
 \eqn\finalmine{ {1 \over 2} \tilde \CR \tilde \CR'' - {1 \over 4} (\tilde \CR')^2
 - 4 \tilde \CR^2 (\tilde \CR+ \hat u) =0 \ .}
Substituting $\hat u= \tilde u + \tilde z/2$ we end up with
 \eqn\final{\eqalign{
 &8 \tilde u \tilde \CR^2 -{1 \over 2} \tilde \CR \tilde \CR'' + {1 \over 4} (\tilde \CR')^2
  =0 \ ,\cr
 & \tilde \CR = \tilde u - \tilde z/2
\ ,\cr & F'' =  4 \tilde u\ .
 }}

\subsec{ $ q>0$}

For $q>0$, we start again with the left-hand sides in \recrel\ and
integrate by parts. The boundary term vanishes, but there is an
extra term where the derivative acts on the factor of $\lambda^q$.
We again call them $\Omega_n$, $\tilde \Omega_n$:
 \eqn\qomega{\eqalign{
 {N \over \g q  } \Omega_n =& h_n^{-1} \int d\mu { 1 \over y}
 P_n(y)^2 = h_n^{-1}\int d\mu { 1 \over y }
 (P_n(y)-P_n(0) )P_n(y)  + h_n^{-1}P_n(0) \int d\mu
 { 1 \over y } P_n(y) \cr
  =&  h_n^{-1} P_n(0)^2 \int { d\mu  \over y}  + P_n(0)
 P_n'(0) h_0/h_n \ ,\cr
  { N\over \g q} \tilde \Omega_n =& h_{n-1}^{-1} P_n(0) P_{n-1}(0)
  \int d\mu { 1 \over y}+ P_{n-1}(0) P_n'(0) h_0/h_{n-1}  \cr
  =& h_{n-1}^{-1} P_n(0) P_{n-1}(0) \int d\mu { 1 \over y}
 + P_{n}(0) P_{n-1}'(0) h_0/h_{n-1} + 1\ .
 }}
Since the measure includes a factor of $y^q$, all these integrals
converge.  Note that in the formal limit $q\to 0$ we recover the
values for $\Omega $ and $\tilde \Omega $ of \omomt. Using
\qomega\ (and the two forms of $\tilde \Omega$), we find the
generalization of \eqnsforomega
 \eqn\eqnomegaq{\eqalign{
 &\tilde \Omega_{n} + \tilde \Omega_{n+1} = - s_n \Omega_n + { \g q
 \over N} \ ,\cr & r_n \Omega_n \Omega_{n-1} = \tilde \Omega_n^2 - { q
 \g \over N} \tilde \Omega_n \ .}}

The corrections due to $q$ are of order $1/N$; therefore, they do
not contribute in the planar limit.  However, they have to be kept
in the double scaling limit.  Repeating the derivation of the
differential equation \finalmine\ we find
 \eqn\equaq{\eqalign{
 & 16\,\tilde \CR^3 + 16\,\tilde \CR^2\,\hat u +
  (\tilde \CR')^2 - 2\,\tilde \CR\,\tilde \CR'' =q^2\ , \cr
   & \tilde \CR = \hat u - \tilde z \ . }}
Substituting $\hat u= \tilde u + \tilde z/2$ we end up with
 \eqn\qfinal{\eqalign{
 &32\,\tilde \CR^2\,\tilde u +
  (\tilde \CR')^2 - 2\,\tilde \CR\, \tilde \CR'' =q^2  \ ,\cr
 & \tilde \CR = \tilde u - \tilde z/2\ , \cr & F'' =  4 \tilde u\ .
 }}
Equations \stringfi, \heat\ may be brought to this form
by defining
\eqn\utu{ u= 2^{5/3} \tilde u\ ,\qquad  z = 2^{2/3}
\tilde z\ ,}
from which it follows that $\CR = 2^{5/3} \tilde
\CR.$
No rescaling of $q$ is needed. This completes the derivation for $k=1$.

Note that \qomega\ also implies that $\Omega \geq 0$, and that it
approaches zero as $\tilde z \to + \infty $. The equations
\eqnsforomega\ imply that in the scaling limit, where $s_n$ is
close to one, $\tilde \Omega$ is negative. Then \scal\ implies
that $\tilde \CR$ is positive and that $\tilde \CR \to 0$ as
$\tilde z \to \infty $. It can also be seen from the second sphere
solution above that $\tilde \CR \sim - \tilde z/2 $ for $\tilde z
\to - \infty $, which is indeed a property that we get from
\stringq\ once we impose that $u \to 0$ as $\tilde z \to -
\infty$.

\subsec{The Miura transformation}

In this subsection we consider the ZS hierarchy for $g=0 = \omega$,
so that $f=r$. In this case only the $F_l$ in \oper\ with
$l$ even, $l=2k$,
are nonzero. These are the terms in the mKdV hierarchy. There is
an interesting relation between these $F_{2k}(f,g=0) =
R_{2k}(r,\omega=0)$
and the Gelfand-Dikii
Polynomials $Q_k(u)$ \gelfand .
The functions $F_{2k}$ can be defined in terms of matrix elements of
the  operator
\eqn\operator{\eqalign{
 {\cal O} =& { 1 \over (d_x + f J_1 - \zeta J_3)} =
 ( d_x - f J_1 - \zeta J_3 ) { 1 \over  (d_x + f J_1 - \zeta J_3)
( d_x - f J_1 - \zeta J_3 ) } =
\cr
=& ( d_x - f J_1 - \zeta J_3 )
{ 1 \over  d_x^2 - f' J_1 - {f^2 \over 4}  - \zeta^2/4 }
}}
We can derive  a similar relation where we write the numerator on the
right.
Taking the $\langle x| \cdots |x\rangle $ matrix elements of these
operators, extracting the piece proportional to $1$, which should
vanish, and the piece proportional to $J_1$, which should equal
$- \sum F_{2k} \zeta^{-2 k -1}$,  we find that
\eqn\foundrel{
F_{2k} =   2^{2 k+1} (d_x +f) Q_{k}(u_-) =
-2^{2 k+1} (d_x -f) Q_{k}(u_+)
}
where
\eqn\defofu{
u_{\pm} = {f^2 \over 4} \pm {f' \over 2}
}
These relations can be viewed as arising from supersymmetric
quantum mechanics, where the operator ${\cal O}$ is the resolvent of
the supercharge, and the operator $d_x^2 + u$ is the Hamiltonian.

Let us introduce a derivative ${\cal D}_+ = d_x + f $ and
assume that $u$ in \stringq\ can be written in terms
of $f$ through \defofu\ as $u = u_-$.
Then it is possible to show that
 \stringq\  becomes
\eqn\equation{
q^2= { f \over 2}  \CR {\cal D}_+ \CR - {1 \over 2}
\CR {\cal D}_+^2 \CR +
{ 1 \over 4} ({\cal D}_+
\CR)^2 =
-{ 1 \over 2}  \CR d_x( {\cal D}_+ {\cal R} ) + { 1 \over 4}
({\cal D}_+ \CR)^2
}
Acting with  ${\cal D}_+$ on the second line in \stringq\ we get
\eqn\secondlin{
{\cal D}_+ { \cal R} = \sum_{k\geq 0} (2k+1) \tilde t_{ 2 l} F_{2l} -1
~,~~~~~~\tilde t_{2k} = 2^{-2k-2} t_k }
Note that in \stringq\ we are implicitly saying that $t_0 =-4 x$.
This translates into $\tilde t_0 = - x $. This $x$ dependence of
$t_0$ is the source of the $-1$ in the right hand side of \secondlin .
These relations were found in \refs{\Hollowood,\DalleyBR}.

In conclusion suppose that we find a solution $f$ that solves
\secondlin\ with ${\cal D}_+ \CR  = 2 q $ with $q\geq 0$.
In other words, a solution of the equation
\eqn\mkdvtype{
2 q + 1 = \sum_{k\geq 0} (2k+1) \tilde t_{ 2 k} F_{2k}
}
Then we will also solve \equation .
In \CrnkovicMS\ it was argued that smooth
solutions exist for the mKdV equation with a zero left hand side.
Since \mkdvtype\ follows from a lagrangian that is bounded below\foot{
We have checked up to $k=3$ that all the terms with derivatives
in $\int H_{2k+1} (f,g=0)$ are positive definite. We think it is true
in general.},
if the highest nonzero $\tilde t_{2k}$ has the right sign, then
it is easy to see that a smooth solution should exist that
interpolates between $f \sim x^{ 1/2m}$ for large $x$ and
$f \sim - ( 2 q + 1)/x $  for large negative $x$.
 Note that if
$2 q + 1$ is zero, then $f \to -f$ is a symmetry of the problem. On
the other hand if $2q+1$ is nonzero the minimum of the action
at large $x$ with positive $f$ is the one with lowest energy.

This solution is such that $u$ obeys the physically relevant
boundary conditions $u \to x^{1/m} $ for $x \to + \infty$ and
$u \to 0$ for $x \to \infty$. We should also impose that
 ${\CR} \geq  0 $ and that it goes to zero for large $x$,
as we showed above.
By integrating the equation ${\cal D}_+ R = 2q $
we find that
\eqn\eqnforr{
\CR = 2q e^{ - \int_{x_0}^x f }
\int_{-\infty}^x dy e^{ \int^y_{x_0} f }
}
 For $q\geq0$ and with
$f$ obeying the above boundary conditions we see that $\CR$ has the
requisite properties.
For $q=0$ it was proven in \johnsonflows , that a unique solution of
\stringq\ exists.

We can study the concrete example of $k=1$ in order to see how this
works \refs{\DalleyBR,\LafranceWY}. Defining $u = u_-$ in \defofu ,
we find
\eqn\painlget{
 2q + 1 = - { 1 \over 2} f'' +  { 1 \over 4} f^3 - x f
}
We see that this equation comes from an action bounded below and
that the lowest minimum at large $x $ has $f \sim { 1 \over 2 } x^{1/2}$.

Note that if we had imposed instead
${\cal D}_+ \CR = - 2 |q|$ we would also solve
\equation , but $\CR$ would not obey the right conditions.

This relation between \stringq\ and the equation
\mkdvtype , can also be understood as follows.
When we are dealing with a hermitian two-cut model with a
symmetric potential,
we can set $g=0$ and we can consider symmetric and anti-symmetric
polynomials independently. Namely, the free energy can be
expressed as a sum $F = F_+ + F_-$ where $F_\pm$ represents
the contribution to the integral from even and odd polynomials.
Then we have that $F_\pm'' = { f^2 \over 8} \mp {f' \over 4} $,
where $f$ obeys the same equation for both cases. Of course,
$F'' = f^2/4$  is the equation for the free energy for the
two cut model.
In the hermitian model we can add a logarithmic
potential of the form $ M \log |\lambda| $ \minahan , which
introduces a factor of $|\lambda|^M$ in the measure that defines
the orthogonal polynomials. The resulting equation for $f$ is the
same as in  \eomaaa\ but with
 a constant proportional to $M$ added to the right
hand side\foot{Normalizations in \minahan\ are different,
$f_{Minahan} = f_{here}/2$. } (remember that we have set $g=0$).
Let us consider now the complex matrix model, and look at the
first form of the integral in \compdefa . Then we can take the
even polynomials $P_{2l}$
 of the hermitian matrix
model defined with the measure $d\mu = \lambda^{1+ 2 q} e^{ -{ N
\over \gamma} V(\lambda^2) }$. These polynomials are functions of
$\lambda^2$ and we can rewrite the Vandermonde determinant that
appears in the first line of \compdefa\ in terms of them. This
will lead to an expression for the partition function of the
complex matrix model in terms of the norms of the
 even polynomials of
a two-cut model with a logarithmic potential with
coefficient $M = 1 + 2 q$. This last
problem is precisely the one solved by the equation \painlget ,
with free energy given by $F'' = f^2/8 - f'/4$.

\appendix{E}{The loop equation of the complex matrix model}

For $M$ a complex rectangular $N\times (N+q)$ matrix, with $q$
positive, there are two closely related resolvents
 \eqn\comres{R={1\over N} \Tr {1 \over M^\dagger M -z},
 \qquad\qquad  \tilde R
 = {1\over N}\Tr {1 \over MM^\dagger -z}= R -{q\over N z}.}
Let us derive their loop equations.  We compute
 \eqn\exploo{\int dM dM^\dagger \Tr\left( {\partial
 \over \partial M} M
 {1 \over M^\dagger M -z} \right) e^{-N\Tr V(M^\dagger M)}=0}
which vanishes because it is a total derivative.\foot{ Compared to
\compdef, we absorb $1/\gamma$ into the definition of $V(M^\dagger
M)$.} This reflects the invariance of the integral under the
change of variables $\delta M=M{1\over M^\dagger M-z}$. It is
straightforward to calculate \exploo\
 \eqn\matinl{\eqalign{
 &-N \left\langle \Tr{  M^\dagger M V'(M^\dagger M) \over
 M^\dagger M -z}\right\rangle + (N+q) \left\langle \Tr {1 \over
 M^\dagger M -z}\right\rangle- \left\langle \Tr {M^\dagger M \over
 M^\dagger M -z} \Tr  {1 \over M^\dagger M -z} \right\rangle \cr
 =&-N \left\langle \Tr { M^\dagger M V'(M^\dagger M) -zV'(z) \over
 M^\dagger M -z}\right\rangle - Nz V'(z) \left\langle \Tr{1
 \over M^\dagger M -z}\right\rangle \cr
 &\qquad + q \left\langle \Tr {1 \over M^\dagger M -z}\right\rangle
 -z  \left\langle \Tr  {1 \over (M^\dagger M -z)^2}
 \right\rangle\cr
 =&-zN^2 \left( \left\langle R(z)^2\right\rangle
 +\left(V'(z)-{q\over N z}\right) \left\langle R(z)\right\rangle
 -{ f(z) \over 4 z} \right) }}
with $f(z)=-{4\over N}\left\langle \Tr{ M^\dagger M V'(M^\dagger
M)-zV'(z) \over M^\dagger M -z}\right\rangle $ a polynomial of the
same degree as $V'(z)$.

In the large $N$ limit the term $\left\langle R(z)^2\right\rangle$
factorizes.  We do not neglect the term proportional to $\hat
q={q\over N}$ because we allow for the possibility that $q$ is
going to infinity with finite $\hat q$.  We derive the loop
equation
 \eqn\loopec{R(z)^2 + \left(V'(z) -{\hat q \over z}\right) R(z)-
 {f(z)\over 4z} =0}
where we denoted the expectation value of the operator $R$ by $R$.
As a check, we can substitute $R=\tilde R +{\hat q\over z}$ and
derive the loop equation for $\tilde R$
 \eqn\loopect{ \tilde R(z)^2 + \left(V'(z) +{\hat q \over z}\right)
 \tilde R(z)  - {f(z) -4\hat qV'(z) \over4 z}=0 }
i.e.\ exactly the same as \loopec\ but with $R \to \tilde R$,
$q\to -q$, and the transformation of $f(z)$ which follows from
$M^\dagger M \to MM^\dagger$.

The solution of \loopec\ is
 \eqn\Rsol{2R(z)=-\left(V'(z) +{\hat q \over z}\right) \pm \sqrt{
 \left(V'(z) +{\hat q \over z}\right)^2+ {f(z) \over z}}}
We see that the parameter $z$ in $R(z)$ takes values in a two fold
cover of the complex plane which is the Riemann surface
 \eqn\Rsur{y^2=
 \left(V'(z) +{\hat q \over z}\right)^2+{f(z) \over z},}
or equivalently in terms of $\hat y=yz$
 \eqn\Rsurh{\hat y^2= \left(zV'(z) +\hat q \right)^2+z f(z).}

As $z\to \infty$ on the upper sheet, we have $R \to -{1\over z}$.
This determines
 \eqn\Rsol{2R(P_\pm(z))=-\left(V'(z) +{\hat q \over z}\right)
 \pm \sqrt{
 \left(V'(z) +{\hat q \over z}\right)^2+ {f(z) \over z}}}
where $P_\pm(z)$ denote the points on the upper and lower sheets
which correspond to $z$. Finally, note that $R$ has a pole on the
lower sheet $R\approx -{\hat q\over P_-(z)}$. The other resolvent
has a pole in the upper sheet $\tilde R\approx {\hat q\over
P_+(z)}$.

Let us consider the simplest model with $V(z)={1 \over \g}
(-z+z^2/2)$.
 The
polynomial $\hat y^2$ in \Rsurh\ is of fourth order, describing a
genus one surface.  The polynomial $f(z)$ is $f = { -4 z/\g} +
f_0$. If we are interested in a one-cut model we impose that the
polynomial has a double zero. This determines the constant $f_0$.
 The most general
such polynomial is \eqn\simpmoc{ \hat y^2= { 1 \over \g^2}
(z-a)^2\left[z^2 + 2z (a-1) + (1 - 4 a + 3 a^2 - 4\g - 2 \g \hat
q) \right] } where $a$ obeys the equation \eqn\equfora{ 4 a^3 - 3
a^4 + \g^2 {\hat q}^2 + a^2 (4\g -1 + 2 \g \hat q) =0 \ .} We are
interested in the double scaling limit, where one of the ends of the cut
approaches $z\sim 0$. Let us first set $\hat q=0$. From \equfora\
we see that $a=0$ is always a solution and leads to a cut
extending from $ 1 - \sqrt{ 4\g}$ to $1+ \sqrt{4\g} $. This
solution makes sense as long as $ 4\g <1$, otherwise the cut would
extend to negative values of $z$ which are not allowed physically.

Thus, for $4\g>1$ the solution must have $a \not =0$. We write
$4\g = 1- \delta$ and take the limit of small negative $\delta$
and small $a$. From \equfora\ with $\hat q =0$, we see that we get
a solution with $ 4 a =\delta$. For $\hat q$ nonzero we see that
we need to scale $\hat q$ as $ |\delta|^{3/2}$. Then we find that
\equfora\ simplifies to \eqn\equsimple{ 4 a^3 +{ \hat q^2 \over
16} - \delta a^2 =0 \ ,} and the second derivative of the free
energy is proportional to the position $z$ of the cut closest to
the origin, which is given by $u = - 4 a + \delta $, in the
scaling limit. After we scale $-a\sim \hat q^{2/3} v$, and
$-\delta \sim \hat q^{2/3} t$, then \equsimple\ reduces to
\cubicv.

To summarize, for $\hat q=0$ we have a phase transition between
the regime with $a=0$ and $a \not = 0$. In one phase $\hat y^2$
has a double zero at $z=0$, and in the other phase the double zero
is at negative $z$ while the cut reaches $z=0$.  For $\hat
q\not=0$ there is no such transition, and the cut does not reach
$z=0$.

Let us now discuss the branes of 0A theory. The FZZT branes
with $\eta = \pm 1$ in the 0A are the same as the
FZZT branes with $\eta = \mp 1 $ in the 0B theory (at $\hat c =0$).
We think that the
resolvent \comres\ corresponds to the $FZZT$ brane with
$\eta =-1$\foot{ Or the FZZT with $\eta = +1$ in 0B.}
 and $\mu_B^2
= - z$. Indeed we see that the disk diagram has the expected one
cut when expressed in terms of $z$ \integrexp . On the other hand,
for negative $\mu$, this brane will have an expectation value
similar to the $FZZT$ brane with $\eta = 1$
at positive $\mu$ \diskdia , which
in terms of $z$ has a cut at the origin as expected.

\listrefs

\bye